\documentclass[final,5p,twocolumn,times]{elsarticle}
\usepackage{amsmath,amssymb,amsfonts,bm,latexsym}
\usepackage{fancybox}
\usepackage{fancyvrb}
\usepackage{multicol}
\usepackage{graphicx}
\usepackage{verbatim}
\usepackage{esdiff}
\usepackage{xcolor}
\usepackage{xspace}
\usepackage{caption}
\usepackage{subcaption}
\usepackage{url}

\usepackage[autolanguage]{numprint}

\newcommand{\nek}{\emph{Nektar++}\xspace}
\newcommand{\shp}{spectral/$hp$\xspace}

\usepackage{tikz,forest}
\usepackage[utf8]{inputenc}
\usepackage[british]{babel}
\usepackage{listings}
\usepackage[autolanguage]{numprint}
\usepackage{booktabs}
\usetikzlibrary{matrix,positioning,calc}

\journal{Computer Physics Communications}

\usepackage{microtype}
\usepackage[T1]{fontenc}
\newcommand\Small{\fontsize{9}{9.2}\selectfont}
\newcommand*\LSTfont{\Small\ttfamily\SetTracking{encoding=*}{-60}\lsstyle}

\usepackage{color}
\definecolor{deepblue}{rgb}{0,0,0.5}
\definecolor{deepred}{rgb}{0.6,0,0}
\definecolor{deepgreen}{rgb}{0,0.5,0}

\newcommand\pythonstyle{\lstset{
    language=Python,
    basewidth=0.5em,
    basicstyle=\LSTfont,
    morekeywords={self,as},             
    keywordstyle=\color{deepblue},
    stringstyle=\color{deepgreen},
    commentstyle=\color{deepgreen},
    frame=tb,                         
    showstringspaces=false            %
}}

\lstnewenvironment{python}[1][]
{
\pythonstyle
\lstset{#1}
}
{}

\definecolor{gray}{rgb}{0.4,0.4,0.4}
\definecolor{darkblue}{rgb}{0.0,0.0,0.6}
\definecolor{cyan}{rgb}{0.0,0.6,0.6}
\definecolor{maroon}{rgb}{0.5,0.0,0.0}
\definecolor{darkgreen}{rgb}{0.0,0.5,0.0}
\lstdefinelanguage{XML}
{
  basicstyle=\ttfamily\footnotesize,
  morestring=[b]",
  moredelim=[s][\bfseries\color{maroon}]{<}{\ },
  moredelim=[s][\bfseries\color{maroon}]{</}{>},
  moredelim=[l][\bfseries\color{maroon}]{/>},
  moredelim=[l][\bfseries\color{maroon}]{>},
  morecomment=[s]{<?}{?>},
  morecomment=[s]{<!--}{-->},
  commentstyle=\color{gray},
  stringstyle=\color{orange},
  identifierstyle=\color{darkblue},
  showstringspaces=false
}
\lstdefinestyle{XMLStyle}{
  language=XML,
  basicstyle=\ttfamily\footnotesize,
  numbers=left,
  numberstyle=\tiny,
  numbersep=3pt,
  frame=,
  columns=fullflexible,
  backgroundcolor=\color{black!05},
  linewidth=\linewidth,
  xleftmargin=0.05\linewidth,
  keepspaces=true
}

\begin{document}

\begin{frontmatter}
  \title{\nek: enhancing the capability and application of high-fidelity
    spectral/$hp$ element methods}

  \author{David~Moxey\fnref{EXE}}
  \author{Chris~D.~Cantwell\fnref{IMP}}
  \author{Yan~Bao\fnref{SJTU}}
  \author{Andrea~Cassinelli\fnref{IMP}}
  \author{Giacomo~Castiglioni\fnref{IMP}}
  \author{Sehun~Chun\fnref{YON}}
  \author{Emilia~Juda\fnref{IMP}}
  \author{Ehsan~Kazemi\fnref{YON}}
  \author{Kilian~Lackhove\fnref{TU}}
  \author{Julian~Marcon\fnref{IMP}}
  \author{Gianmarco~Mengaldo\fnref{CAL}}
  \author{Douglas~Serson\fnref{IMP}}
  \author{Michael~Turner\fnref{IMP}}
  \author{Hui~Xu\fnref{SJTUAERO,IMP}}
  \author{Joaquim~Peir\'o\fnref{IMP}}
  \author{Robert~M.~Kirby\fnref{SCI}}
  \author{Spencer~J.~Sherwin\fnref{IMP}}

  \fntext[EXE]{College of Engineering, Mathematics and Physical Sciences,
    University of Exeter, United Kingdom}
  \fntext[IMP]{Department of Aeronautics, Imperial College London, United
    Kingdom}
  \fntext[SJTU]{Department of Civil Engineering, Shanghai Jiao Tong University,
    Shanghai, China}
  \fntext[YON]{Underwood International College, Yonsei University, South Korea}
  \fntext[SJTUAERO]{School of Aeronautics and Astronautics, Shanghai Jiao Tong
    University, Shanghai, China}
  \fntext[TU]{Department of Energy and Power Plant Technology, Technische
    Universit\"at Darmstadt, Germany}
  \fntext[CAL]{Division of Engineering and Applied Science. California
    Institute of Technology}
  \fntext[SCI]{Scientific Computing and Imaging Institute, University of Utah,
    USA}

  \begin{abstract}
    

    \nek is an open-source framework that provides a flexible, high-performance
    and scalable platform for the development of solvers for partial
    differential equations using the high-order spectral/$hp$ element method. In
    particular, \nek aims to overcome the complex implementation challenges that
    are often associated with high-order methods, thereby allowing them to be
    more readily used in a wide range of application areas.
    In this paper, we present the algorithmic, implementation and application
    developments associated with our {\nek} version 5.0 release.  We describe
    some of the key software and performance developments, including our
    strategies on parallel I/O, on {\em in situ} processing, the use of
    collective operations for exploiting current and emerging hardware, and
    interfaces to enable multi-solver coupling.  Furthermore, we provide details
    on a newly developed Python interface that enables a more rapid introduction
    for new users unfamiliar with spectral/$hp$ element methods, C++ and/or
    \nek. This release also incorporates a number of numerical method
    developments -- in particular: the method of moving frames (MMF), which
    provides an additional approach for the simulation of equations on embedded
    curvilinear manifolds and domains; a means of handling spatially variable
    polynomial order; and a novel technique for quasi-3D simulations (which
    combine a 2D spectral element and 1D Fourier spectral method) to permit
    spatially-varying perturbations to the geometry in the homogeneous
    direction. Finally, we demonstrate the new application-level features
    provided in this release, namely: a facility for generating high-order
    curvilinear meshes called \emph{NekMesh}; a novel new \emph{AcousticSolver}
    for aeroacoustic problems; our development of a `thick' strip model for the
    modelling of fluid-structure interaction (FSI) problems in the context of
    vortex-induced vibrations (VIV).  We conclude by commenting on some lessons
    learned and by discussing some directions for future code development and
    expansion.

\medskip
\noindent{\bf PROGRAM SUMMARY}

\begin{small}
\noindent
{\em Manuscript Title:} \nek: enhancing the capability and application of high-fidelity
    spectral/$hp$ element methods \\
{\em Authors:} D.~Moxey, C.~D.~Cantwell, Y.~Bao, A.~Cassinelli, G.~Castiglioni, S.~Chun, E.~Juda, E.~Kazemi, K.~Lackhove, J.~Marcon, G.~Mengaldo, D.~Serson, M.~Turner, H.~Xu, J.~Peir\'o, R.~M.~Kirby, S.~J.~Sherwin \\
{\em Program Title:} Nektar++                                 \\
{\em Journal Reference:}                                      \\
{\em Catalogue identifier:}                                   \\
{\em Licensing provisions:} MIT                               \\
{\em Programming language:} C++                               \\
{\em Computer:} Any PC workstation or cluster                 \\
{\em Operating system:} Linux/UNIX, macOS, Microsoft Windows   \\
{\em RAM:} 512 MB+                                             \\
{\em Number of processors used:} 1-1024+                      \\
{\em Supplementary material:}                                 
{\em Classification:} 12 Gases and Fluids                     \\
{\em External routines/libraries:} Boost, METIS, FFTW, MPI, Scotch,
PETSc, TinyXML, HDF5, OpenCASCADE, CWIPI \\
{\em Nature of problem:} The \nek framework is designed to enable the
discretisation and solution of time-independent or time-dependent partial
differential equations.\\
{\em Solution method:} \shp{} element method\\
{\em Reasons for the new version:}*\\
{\em Summary of revisions:}*\\
{\em Restrictions:}\\
{\em Unusual features:}\\
{\em Additional comments:}\\
{\em Running time:} The tests provided take a few minutes to run. Runtime in
general depends on mesh size and total integration time.\\

\end{small}
\end{abstract}

  
\end{frontmatter}



\section{Introduction}
\label{sec:introduction}

High-order finite element methods are becoming increasingly popular in both
academia and industry, as scientists and technological innovators strive to
increase the fidelity and accuracy of their simulations whilst retaining
computational efficiency. The spectral/$hp$ element method in particular, which
combines the geometric flexibility of classical low-order finite element methods
with the attractive convergence properties of spectral discretisations, can
yield a number of advantages in this regard. From a numerical analysis
perspective, their diffusion and dispersion characteristics mean that they are
ideally suited to applications such as fluid dynamics, where flow structures
must be convected across long time- and length-scales without suffering from
artificial
dissipation~\cite{moura2017eddy,moura2017setting,mengaldo2018spatial_a,mengaldo2018spatial_b,fernandez2019non}
High-order methods are also less computationally costly than traditional
low-order numerical schemes for a given number of degrees of freedom, owing to
their ability to exploit a more locally compact and dense elemental operators
when compared to sparse low-order
discretisations~\cite{VoShKi10,CaShKiKe11a,CaShKiKe11b}. In addition, high-order
methods in various formulations can be seen to encapsulate other existing
methods, such as finite volume, finite difference (e.g. summation-by-parts
finite difference~\cite{gassner2013skew}), finite element, and flux
reconstruction
approaches~\cite{mengaldo2016connections,mengaldo2015discontinuous}. All of
these features make the spectral/$hp$ element method an extremely attractive
tool to practitioners.

As the name suggests, the spectral/$hp$ element method relies on the tesselation
of a computational domain into a collection of elements of arbitrary size that
form a mesh, where each element is equipped with a polynomial expansion of
arbitrary and potentially spatially-variable order~\cite{KaSh05}. Within this
definition, we include continuous Galerkin (CG) and discontinuous Galerkin (DG)
methods, along with their variants. High-order methods have been historically
seen as complex to implement, and their adoption has been consequently limited
to academic groups and numerical analysts. This mantra is rapidly being removed
thanks to the development of open-source numerical libraries that facilitate the
implementation of new high-fidelity solvers for the solution of problems arising
across a broad spectrum of research areas in engineering, biomedicine, numerical
weather and climate prediction, and economics. An additional challenge in the
use of high-order methods, particularly for problems involving complex
geometries, is the generation of a curvilinear mesh that conforms to the
underlying curves and surfaces. However, advances in curvilinear mesh generation
(such as~\cite{turner2018curvilinear}), combined with open-source efforts to
increase their prevalence, mean that simulations across extremely complex
geometries are now possible.

\nek is a project initiated in 2005 (the first commit to an online repository
was made on 4$^{\mathrm{th}}$ May 2006), with the aim of facilitating the
development of high-fidelity computationally-efficient, and scalable spectral
element solvers, thereby helping close the gap between application-domain
experts (or users), and numerical-method experts (or developers). Along with
\nek, other packages that implement such high-order methods have been developed
in the past several years. Nek5000, developed at Argonne National Laboratory,
implements CG discretizations mainly for solving incompressible and low-Mach
number flow problems on hexahedral meshes using the classical spectral element
method of collocated Lagrange
interpolants~\cite{Nek5000}. Semtex~\cite{BlSh04,blackburn-2019} is a fluid
dynamics code that also uses the classical spectral element method in two
dimensions, with the use of a 1D pseudospectral Fourier expansion for three
dimensional problems that incorporate a homogeneous component of
geometry. \nek also supports this joint discretisation in Cartesian
coordinates; however Semtex also includes support for cylindrical coordinate
systems, where the Fourier modes are used in the azimuthal direction, which
broadens the range of geometries that can be considered in this
setting. deal.II~\cite{BaHaKa07} is a more generic finite element framework,
which likewise restricts its element choices to quadrilaterals and hexahedra,
but permits the solution of a wide array of problems, alongside the use of
adaptive mesh refinement.  Flexi~\cite{hindenlang2012explicit} and its spinoff
Fluxo~\cite{fluxo_SplitForm}, developed at the University of Stuttgart and at
the University of Cologne, implement discontinuous Galerkin methods for flow
problems on hexahedral meshes. GNuME, and its NUMO and NUMA components developed
at the Naval Postgraduate School, implement both continuous and discontinuous
Galerkin methods mainly for weather and climate prediction
purposes~\cite{giraldo2008study,abdi2016efficient}. PyFR~\cite{WiFaVi14},
developed at Imperial College London, implements the flux reconstruction
approach~\cite{huynh2007flux} which shares various numerical properties with DG
in particular~\cite{mengaldo2016connections,allaneau2011connections}. DUNE
implements a DG formulation, among a wide variety of other numerical methods
such as finite difference and finite volume methods~\cite{DeKlNoOh11}.

\nek is a continuation of an earlier code \emph{Nektar}, itself developed at
Brown University originally using the C programming language, with some parts
extended to use C++. \nek is instead written using the C++ language, and greatly
exploits its object-oriented capabilities. The aim of \nek is to encapsulate
many of the high-order discretisation strategies mentioned previously, in a
readily accessible framework. The current release includes both CG and DG
methods and, arguably, its distinguishing feature is its inherent support for
complex geometries through various unstructured high-order element types; namely
hexahedra, tetrahedra, prisms and pyramids for three-dimensional problems, and
quadrilaterals and triangles for two-dimensional problems. Both CG and DG can be
used on meshes that contain different element shapes (also referred to as hybrid
meshes), and allow for curvilinear element boundaries in proximity of curved
geometrical shapes. Along with these spatial discretizations, \nek supports
so-called quasi-3D simulations in a manner similar to Semtex, where a
geometrically complex 2D spectral element mesh is combined with a classical 1D
Fourier expansion in a third, geometrically homogeneous, direction. This mixed
formulation can significantly enhance computational efficiency for problems of
the appropriate geometry~\cite{BlSh04} and \nek supports a number of different
parallelisation strategies for this approach~\cite{bolis2016adaptable}. The time
discretization is achieved using a general linear method formulation for the
encapsulation of implicit, explicit and mixed implicit-explicit timestepping
schemes~\cite{VoEsBoChKi11}. While the main purpose of the library is to create
an environment under which users can develop novel solvers for the applications
of their interest, \nek already includes fully-fledged solvers for the solution
of several common systems, including fluid flows governed either by the
incompressible or compressible Navier-Stokes and Euler equations;
advection-diffusion-reaction problems, including on a manifold, with specific
applications to cardiac electrophysiology~\cite{CaYaKiPeSh14}. One of the main
shortcomings of the spectral/$hp$ element method is related to a perceived lack
of robustness, arising from low dissipative properties, which can be a
significant challenge for industrial applications. \nek implements several
techniques to address this problem, namely efficient dealiasing
techniques~\cite{mengaldo2015dealiasing,winters2018comparative} and spectral
vanishing viscosity~\cite{kirby2006stabilisation}, that have proved invaluable
for particularly challenging applications~\cite{lombard2015implicit}.

The scope of this review is to highlight the substantial number of new
developments in \nek since the last publication related to the software,
released in 2015 and coinciding with the version 4 release of
\nek~\cite{nektarpp2015}. Since this release, over \numprint{7000} commits have
been added to the main code for the version 5 documented here, with a key focus
on expanding the capability of the code to provide efficient high-fidelity
simulations for challenging problems in various scientific and engineering
fields. To this end, the paper is organized as follows. After a brief review of
the formulation in Section~\ref{sec:methods}, in Section~\ref{sec:software} we
present our software and performance developments. This includes our strategies
on parallel I/O; {\em in situ} processing; the use of collective linear algebra
operations for exploiting current and emerging hardware; and interfaces for
multi-solver coupling to enable multi-physics simulations using
\nek. Furthermore, we provide details on our new Python interfaces that enable
more rapid on-boarding of new users unfamiliar with either spectral/$hp$ element
methods, C++ or {\nek}.  In Section~\ref{sec:numerics}, we then present recent
numerical method developments -- in particular, the method of moving frames
(MMF); recently added support for spatially-variable polynomial order for
$p$-adaptive simulations; and new ways of incorporating global mappings to
enable spatially variable quasi-3D approaches.  In
Section~\ref{sec:applications}, we then demonstrate some of the new features
provided in our new release, namely: our new facility for generating high-order
(curvilinear) meshes called \emph{NekMesh}; a new \emph{AcousticSolver} for
aeroacoustic problems; and our development of a `thick' strip model for enabling
the solution of fluid-structure interaction (FSI) problems in the context of
vortex-induced vibrations (VIV).  We conclude in Section~\ref{sec:conclusions}
by commenting on some lessons learned and by discussing some directions for
future code development and expansion.

\paragraph{\bf Contributors} \nek{} has been developed across more than a
decade, and we would like to explicitly acknowledge the many people who have
made contributions to the specific application codes distributed with the
libraries. In addition to the coauthors of the previous
publication~\cite{nektarpp2015} we would like to explicitly thank the following
for their contributions:
\begin{itemize}
  \item Dr.~Rheeda Ali (Department of Biomedical Engineering, Johns Hopkins
  University, USA) and Dr.~Caroline Roney (Biomedical Engineering Department,
  King's College London, UK) for their work on the cardiac electrophysiology
  solver;
  \item Dr.~Michael Barbour and Dr.~Kurt Sansom (Department of Mechnical
  Engineering, University of Washington, USA) for developments related to
  biological flows;
  \item Mr.~Filipe Buscariolo (Department of Aeronautics, Imperial College
  London, UK) for contributions to the incompressible Navier-Stokes solver;
  \item Dr.~Jeremy Cohen (Department of Aeronautics, Imperial College London,
  UK) for work relating to cloud deployment and the Nekkloud interface;
  \item Mr.~Ryan Denny (Department of Aeronautics, Imperial College London,
  UK) for enhancements in the 1D arterial pulse wave model;
  \item Mr.~Jan Eichst\"{a}dt (Department of Aeronautics, Imperial College
  London, UK) for initial investigations towards using many-core and GPU
  platforms;
  \item Dr.~Stanis\l{}aw Gepner (Faculty of Power and Aeronautical Engineering,
  Warsaw University of Technology, Poland) for enhancements in the Navier-Stokes
  linear stability solver;
  \item Mr. Dav de St. Germain (SCI Institute, University of Utah, USA) for
  enhancements of timestepping schemes;
  \item Mr. Ashok Jallepalli (SCI Institute, University of Utah, USA) for
  initial efforts on the integration of SIAC filters into post-processing
  routines;
  \item Prof. Dr. Johannes Janicka (Department of Energy and Power Plant
  Technology), Technische Universit\"at Darmstadt, Germany), for support and
  development of the acoustic solver and solver coupling;
  \item Mr.~Edward Laughton (College of Engineering, Mathematics and Physical
  Sciences, University of Exeter, UK) for testing enhancements and initial
  efforts on non-conformal grids;
  \item Dr.~Rodrigo Moura (Divis\~{a}o de Engenharia Aeron\'autica, Instituto
  Tecnol\'ogico de Aeron\'autica, Brasil) for numerical developments related to
  spectral vanishing viscosity stabilisation;
  \item Dr.~Rupert Nash and Dr.~Michael Bareford (EPCC, University of Edinburgh,
  UK) for their work on parallel I/O; and
  \item Mr.~Zhenguo Yan and Mr.~Yu Pan (Department of Aeronautics, Imperial
  College London, UK) for development of the compressible flow solver;
\end{itemize}
%
%
%
%

\section{Methods}
\label{sec:methods}


In this first section, we outline the mathematical framework that underpins
\nek, as originally presented in~\cite{nektarpp2015,jhd}. \nek supports a
variety of spatial discretisation choices, primarily based around the continuous
and discontinuous Galerkin methods (CG and DG). However, in the majority of
cases CG and DG use the same generic numerical infrastructure. Here we therefore
present a brief overview and refer the reader to~\cite{KaSh05} for further
details, which contains a comprehensive description of the method and its
corresponding implementation choices. In the text below we also highlight
appropriate chapters and sections from~\cite{KaSh05} for the material being
discussed.

The broad goal of \nek is to provide a framework for the numerical solution of
partial differential equations (PDEs) of the form $\mathcal{L}u = 0$ on a domain
$\Omega$, which may be geometrically complex, for some solution
$u$. Practically, in order to carry out a spatial discretisation of the PDE
problem, $\Omega$ needs to be partitioned into a finite number of
$d$-dimensional non-overlapping elements $\Omega_e$, where in \nek we support
$1\leq d\leq 3$, such that the collection of all elements recovers the original
region ($\Omega = \bigcup \Omega_e$) and for $e_1\neq e_2$,
$\Omega_{e_1} \cap \Omega_{e_2} = \partial\Omega_{e_1e_2}$ is an empty set or an
interface of dimension $\bar{d}<d$. The domain may be embedded in a space of
equal or higher dimension, $\hat{d} \geq d$, as described
in~\cite{CaYaKiPeSh14}. One of the distinguishing features of \nek is that it
supports a wide variety of elemental shapes: namely segments in one
dimension~\cite[\S2]{KaSh05}; triangles and quadrilaterals in two dimensions,
and; tetrahedra, pyramids, prisms and hexahedra in three
dimensions~\cite[\S3~and~\S4]{KaSh05}. This makes it broadly suitable for the
solution of problems in complex domains, in which hybrid meshes of multiple
elements are generally required.

\nek supports the solution of PDE systems that are either steady-state or
time-dependent; in the case of time-dependent cases, there is subsequently a
choice to use either explicit, implicit or implicit-explicit timestepping. From
an implementation and formulation perspective, steady-state and implicit-type
problems typically require the efficient solution of a system of linear
equations, whereas explicit-type problems rely on the evaluation of the
spatially discretised mathematical operators. In the following sections, we
briefly outline the support in \nek for these regimes.

\subsection{Implicit-type methods}

In this approach we follow the standard finite element derivation as described
in \cite[\S2.2]{KaSh05}, so that before establishing the spatial
discretisation, the abstract operator form $\mathcal{L}u = 0$ is formulated in
the weak sense alongside appropriate test and trial spaces $\mathcal{V}$ and
$\mathcal{U}$. In general, we require at least a first-order derivative and
select, for example,
$\mathcal{V} = H^1_0(\Omega) := \{ v\in H^1(\Omega) \ |\ v(\partial\Omega) = 0
\}$, where
\[
  H^1(\Omega) := \{v \in L^2(\Omega)\,|\,D^{\alpha}u \in
  L^2(\Omega)\,\forall\,|\alpha|\leq 1 \}.
\]
Following the Galerkin approach, we select $\mathcal{U} = \mathcal{V}$. We note
that where problems involve Dirichlet boundary conditions on a boundary
$\partial\Omega_D\subset\Omega$ of the form
$u|_{\partial\Omega_D}(\bm{x}) = g_D(\bm{x})$, we typically enforce this by
lifting $g_D$ as described in~\cite[\S2.2.3~and~\S4.2.4]{KaSh05}. For illustrative
purposes, we assume that $\mathcal{L}$ is linear and its corresponding weak form
can be expressed as: \emph{find $u \in \mathcal{U}$ such that}
\begin{align}\label{eq:var}
  a(u,v) = \ell (v) \quad \forall v \in \mathcal{U},
\end{align} 
where $a(\cdot, \cdot)$ is a bilinear form and $\ell(\cdot)$ is a linear form.

To numerically solve the problem given in Equation~\ref{eq:var} with the spatial
partition of $\Omega$, we consider solutions in an appropriate
finite-dimensional subspace $\mathcal{U}_N\subset\mathcal{U}$. In a high-order
setting, these spaces correspond to the choice of spatial discretisation on the
mesh. For example, in the discontinuous setting we select
\[
  \mathcal{U}_N = \{ u\in L^2(\Omega) \ |\ u|_{\Omega_e} \in \mathcal{P}_P(\Omega_e) \},
\]
where $\mathcal{P}_P(\Omega_e)$ represents the space of polynomials on
$\Omega_e$ up to total degree $P$, so that functions are permitted to be
discontinuous across elemental boundaries. In the continuous setting, we select
the same space intersected with $C^0(\Omega)$, so that expansions are continuous
across elemental boundaries. The solution to the weak form in
Equation~\eqref{eq:var} can then be cast as: \emph{find
  $u^{\delta} \in \mathcal{U}_N$ such that
\begin{align}
  a(u^{\delta}, v^{\delta}) = \ell(v^{\delta})\quad \forall v^{\delta} \in
  \mathcal{U}_N
\label{e:bilinear}
\end{align}
}%
Assuming that the solution can be represented as
$u^{\delta}(\bm{x}) = \sum_n \hat{u}_n\Phi_n(\bm{x})$, a weighted sum of $N$
trial functions $\Phi_n(\bm{x})\in \mathcal{U}_N$ defined on
$\Omega$~\cite[\S2.1]{KaSh05}, the problem then becomes that of finding the
coefficients $\hat{u}_n$, which in the Galerkin approach translates into the
solution of a system of linear equations.


\begin{figure}
  \begin{center}
    \begin{tikzpicture}[scale=2,>=latex]
      \node[below,align=center] at (0.5,-0.2) {collapsed coordinates\\ $\bm{\eta}\in[-1,1]^2$};
      \draw[->] (0,-0.15) -- (0.5,-0.15) node[right] {$\eta_1$} ;
      \draw[->] (-0.15,0) -- (-0.15,0.5) node[above] {$\eta_2$} ;
      \fill[black!10] (0,0) -- (1,0) -- (1,1) -- (0,1) -- cycle;
      \draw[black!60] (0.25,0) -- (0.25,1);
      \draw[black!60] (0.50,0) -- (0.50,1);
      \draw[black!60] (0.75,0) -- (0.75,1);
      \draw[black!60] (0,0.25) -- (1,0.25);
      \draw[black!60] (0,0.50) -- (1,0.50);
      \draw[black!60] (0,0.75) -- (1,0.75);
      \draw[thick] (0,0) -- (1,0) -- (1,1) -- (0,1) -- cycle;

      \draw[<-] (1.2,0.45) -- node[below]{$\bm{\omega}_{\mathcal{T}}^{-1}$} (1.7,0.45);
      \draw[->] (1.2,0.55) -- node[above]{$\bm{\omega}_{\mathcal{T}}$} (1.7,0.55);

      \begin{scope}[xshift=2cm]
        \node[below,align=center] at (0.5,-0.2) {reference coordinates\\ $\bm{\xi}\in\mathcal{T}$};
        \draw[->] (0,-0.15) -- (0.5,-0.15) node[right] {$\xi_1$} ;
        \draw[->] (-0.15,0) -- (-0.15,0.5) node[above] {$\xi_2$} ;
        \fill[black!10] (0,0) -- (1,0) -- (0,1) -- cycle;
        \draw[black!60] (0.25,0) -- (0,1);
        \draw[black!60] (0.50,0) -- (0,1);
        \draw[black!60] (0.75,0) -- (0,1);
        \draw[black!60] (0,0.25) -- (0.75,0.25);
        \draw[black!60] (0,0.50) -- (0.50,0.50);
        \draw[black!60] (0,0.75) -- (0.25,0.75);
        \draw[thick] (0,0) -- (1,0) -- (0,1) -- cycle;
      \end{scope}

      \begin{scope}[xshift=1cm,yshift=1.3cm]
        \draw[->] (0.9,-0.3) -- node[right,yshift=1ex] {$\bm{\chi}_e(\bm{\xi})$} (0.6,0.15);
        \draw[->] (-.55,0.2) -- (-.25,0.2) node[right] {$x_1$};
        \draw[->] (-.55,0.2) -- (-.55,0.5) node[above] {$x_2$};
        \node at (-.4,0) {$\bm{x}\in\Omega_e$};
        \fill[black!10] (0.2,0) to[bend left]
            node[pos=0.25,inner sep=0pt] (c1) {}
            node[pos=0.5, inner sep=0pt] (c2) {}
            node[pos=0.75,inner sep=0pt] (c3) {}
        (1,0.2) to
            node[pos=0.25,inner sep=0pt] (c4) {}
            node[pos=0.5, inner sep=0pt] (c5) {}
            node[pos=0.75,inner sep=0pt] (c6) {}
        (-0.3,1) to[bend left=12]
            node[pos=0.25,inner sep=0pt] (c7) {}
            node[pos=0.5, inner sep=0pt] (c8) {}
            node[pos=0.75,inner sep=0pt] (c9) {}
        cycle;

        \draw[black!60] ($(c1)+(0,-.2pt)$) to[bend right=8] (-0.3,1);
        \draw[black!60] ($(c2)+(0,-.15pt)$) to[bend right=3] (-0.3,1);
        \draw[black!60] ($(c3)+(0,-.1pt)$) to (-0.3,1);
        \draw[black!60] (c4) to[bend right=18] ($(c9)+(0,-.5pt)$);
        \draw[black!60] (c5) to[bend right=13] ($(c8)+(0,-.3pt)$);
        \draw[black!60] ($(c6)+(.2pt,0)$) to[bend right=5] ($(c7)+(0,-.3pt)$);
        \draw[thick] (0.2,0) to[bend left] (1,0.2) -- (-0.3,1) to[bend left=12] cycle;
      \end{scope}
    \end{tikzpicture}
  \end{center}
  \caption{Coordinate systems and mappings between collapsed coordinates
    $\bm{\eta}$, reference coordinates $\bm{\xi}$ and Cartesian coordinates
    $\bm{x}$ for a high-order triangular element $\Omega_e$.}
  \label{f:duffy}
\end{figure}
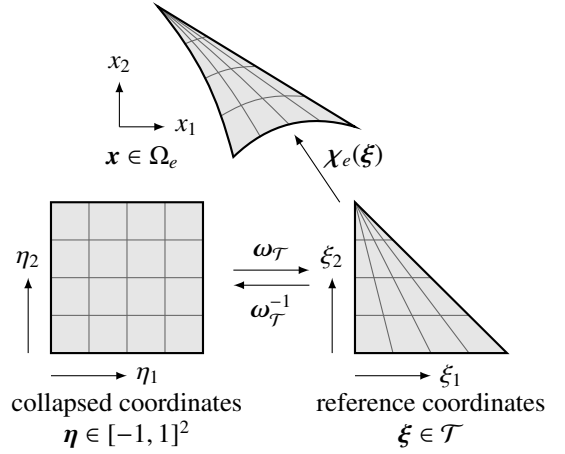

To establish the global basis
$\bm{\Phi}(\Omega) = \{ \Phi_1(\bm{x}), \dots, \Phi_N(\bm{x}) \}$, we first
consider the contributions from each element in the domain.
To simplify the definition of basis functions on each element, we follow the
standard approach described in~\cite[\S2.3.1.2~and~\S4.1.3]{KaSh05} where
$\Omega_e$ is mapped from a standard reference space
$\mathcal{E} \subset [-1,1]^d$ by a parametric mapping
$\bm{\chi}_e: \mathcal{E} \to \Omega_e$, so that
$\bm{x} = \bm{\chi}_e(\bm{\xi})$. Here, $\mathcal{E}$ is one of the supported
region shapes in Table~\ref{t:refregions} and $\bm{\xi}$ are $d$-dimensional
coordinates representing positions in a reference element, distinguishing them
from $\bm{x}$ which are $\hat{d}$-dimensional coordinates in the Cartesian
coordinate space.
On triangular, tetrahedral, prismatic and pyramid elements, one or more of the
coordinate directions of a quadrilateral or hexahedral region are collapsed to
form the appropriate reference shape, creating one or more singular vertices
within these regions~\cite{Du91,ShKa95}. Operations, such as calculating
derivatives, map the tensor-product coordinate system to these shapes through
Duffy transformations~\cite{Du82} --- for example,
$\bm{\omega}_{\mathcal{T}}: \mathcal{T} \rightarrow \mathcal{Q}$ maps the
triangular region $\mathcal{T}$ to the quadrilateral region $\mathcal{Q}$ --- to
allow these methods to be well-defined. The relationship between these
coordinates is depicted in Figure~\ref{f:duffy}. Note that the singularity in
the inverse mapping $\bm{\omega}^{-1}_{\mathcal{T}}$ does not affect convergence
order and can be mitigated in practice by adopting an alternative choice of
quadrature, such as Gauss-Radau points, in order to omit the collapsed
vertices~\cite[\S4.1.1]{KaSh05}.

The mapping $\bm{\chi}_e$ need not necessarily exhibit a constant Jacobian, so
that the resulting element is deformed as shown in Figure~\ref{f:duffy}. \nek
represents the curvature of these elements by taking a sub- or iso-parametric
mapping for $\bm{\chi}_e$, so that the curvature is defined using at least a
subset of the basis functions used to represent the solution
field~\cite[\S4.3.5]{KaSh05}. The ability to use such elements in high-order
simulations is critical in the simulation of complex geometries, as without
curvilinear elements, one could not accurately represent the underlying curves
and surfaces of the geometry, as demonstrated in~\cite{bassi1997high}. The
generation of meshes involving curved elements is, however, a challenging
problem. Our efforts to generate such meshes, as well as to adapt linear meshes
for high-order simulations, are implemented in the \emph{NekMesh} generator tool
described in Section~\ref{sec:nekmesh}, as well as a number of recent
publications (e.g.~\cite{turner2018curvilinear,Marcon2019}).

\begin{table*}
\begin{center}
\begin{tabular}{lll}
\toprule
Name & Class & Domain definition \\
\midrule
Segment         & \texttt{StdSeg}   & $\mathcal{S} = \{ \xi_1 \in [-1,1] \}$ \\
Quadrilateral   & \texttt{StdQuad}  & $\mathcal{Q} = \{ \bm{\xi} \in [-1,1]^2 \}$ \\
Triangle        & \texttt{StdTri}   & $\mathcal{T} = \{ \bm{\xi} \in [-1,1]^2 \, |\, \xi_1 + \xi_2 \leq 0 \}$ \\
Hexahedron      & \texttt{StdHex}   & $\mathcal{H} = \{ \bm{\xi} \in [-1,1]^3 \}$ \\
Prism           & \texttt{StdPrism} & $\mathcal{R} = \{ \bm{\xi} \in [-1,1]^3 \, |\, \xi_1 \leq 1, \xi_2 + \xi_3 \leq 0 \}$ \\
Pyramid         & \texttt{StdPyr}   & $\mathcal{P} = \{ \bm{\xi} \in [-1,1]^3 \, |\, \xi_1 + \xi_3 \leq 0, \xi_2 + \xi_3 \leq 0 \}$ \\
Tetrahedron     & \texttt{StdTet}   & $\mathcal{A} = \{ \bm{\xi} \in [-1,1]^3 \, |\, \xi_1 + \xi_2 + \xi_3 \leq -1 \}$ \\
\bottomrule
\end{tabular}
\end{center}
\caption{List of supported elemental reference regions.}
\label{t:refregions}
\end{table*}

With the mapping $\bm{\chi}_e$ and the transformation
$\bm{\omega}_{\mathcal{T}}$ the discrete approximation $u^{\delta}$ to the
solution $u$ on a single element $\Omega_e$ can then be expressed as
\begin{align*}
  u^{\delta}(\bm{x}) = \sum_n \hat{u}_n
  \phi_n\left(\left[\bm{\chi}_e\right]^{-1}(\bm{x})\right)
\end{align*}
where $\phi_n$ form a basis of $\mathcal{P}_P(\mathcal{E})$; i.e. a local
polynomial basis need only be constructed on each reference element. A
one-dimensional order-$P$ basis is a set of polynomials $\phi_p(\bm{\xi})$,
$0\leq p\leq P$, defined on the reference segment, $\mathcal{S}$.
The choice of basis is usually made based on its mathematical or numerical
properties and may be modal or nodal in nature~\cite[\S2.3]{KaSh05}. For two-
and three-dimensional regions, a tensorial basis may be employed, where the
polynomial space is constructed as the tensor-product of one-dimensional bases
on segments, quadrilaterals or hexahedral regions. In \shp element methods, a
common choice is to use a modified hierarchical Legendre basis (a `bubble'-like
polynomial basis sometimes referred to as the `modal basis'), given
in~\cite[\S3.2.3]{KaSh05} as a function of one variable by
\begin{align*}
\phi_p(\xi) = \left\{\begin{array}{ll}
    \frac{1-\xi}{2} & p = 0,\\
    \left(\frac{1-\xi}{2}\right)\left(\frac{1+\xi}{2}\right)P^{1,1}_{p-1}(\xi) & 0 < p < P,\\
    \frac{1+\xi}{2} & p = P,
\end{array}\right.
\end{align*}
which supports boundary-interior decomposition and therefore improves numerical
efficiency when solving the globally assembled system. Equivalently,
$\phi_p(\xi)$ could also be defined by the Lagrange polynomials through the
Gauss-Lobatto-Legendre quadrature points which would lead to a traditional
spectral element method~\cite[\S2.3.4]{KaSh05}. In higher dimensions, a tensor
product of either basis can be used on quadrilateral and hexahedral elements
respectively. On the other hand, the use of a collapsed coordinate system also
permits the use of a tensor product modal basis for the triangle, tetrahedron,
prism and pyramid, which when combined with tensor contraction techniques can
yield performance improvements. This aspect is considered further in
Section~\ref{sec:collections} and~\cite{moxey-2016b,moxey-2019b}.

Elemental contributions to the solution may be assembled to form a global
solution through an assembly operator~\cite[\S2.3.1~and~\S4.2.1]{KaSh05}. In a
continuous Galerkin setting, the assembly operator sums contributions from
neighbouring elements to enforce the $C^0$-continuity requirement. In a
discontinuous Galerkin formulation, such mappings transfer flux values from the
element interfaces into the global solution vector. For elliptic operators, \nek
has a wide range of implementation choices available to improve computational
performance. A common choice is the use of a (possibly multi-level) static
condensation of the assembled system~\cite[\S4.1.6~and~\S4.2.3]{KaSh05}, where a
global system is formed only on elemental boundary degrees of freedom. This is
supported both for the classical continuous framework, as well as in the DG
method. In the latter, this gives rise to the hybridizable discontinuous
Galerkin (HDG) approach~\cite{yakovlev-2016}, in which a global system is solved
on the trace or skeleton of the overall mesh.

\subsection{Explicit-type methods}

\nek has extensive support for the solution of problems in a time-explicit
manner, which requires the evaluation of discretised spatial operators alongside
projection into the appropriate space. As the construction of the implicit
operators requires these same operator evaluations, most of the formulation
previously discussed directly translates to this approach. We do note however
that there is a particular focus on the discontinous Galerkin method as shown
in~\cite[\S6.2]{KaSh05} for multi-dimensional hyperbolic systems of the form
\[
  \diff{{\bm{u}}}{t} + \nabla\cdot\bm{F}(\bm{u}) = \bm{G}(\bm{u}).
\]
This includes the acoustic perturbation equations that we discuss in
Section~\ref{sec:acoustic} and the compressible Navier-Stokes system used for
aerodynamics simulations in Section~\ref{sec:aero}. In this setting, on a single
element, and further assuming $\bm{G}$ is zero for simplicity of presentation as
in~\cite[\S6.2.2]{KaSh05}, we multiply the above equation by an appropriate test
function $v\in\mathcal{U}$ and integrate by parts to obtain
\[
  \int_{\Omega_e} \diff{{\bm{u}}}{t} v \,\mathrm{d}\bm{x} +
  \int_{\partial\Omega_e} v \tilde{\bm{f}}^e(\bm{u}^{-}, \bm{u}^{+}) \cdot
  \bm{n}\,\mathrm{d}s - \int_{\Omega_e} \nabla v \cdot\bm{F}(\bm{u})
  \,\mathrm{d}\bm{x} = 0.
\]
In the above, $\tilde{\bm{f}}^e(\bm{u}^{-}, \bm{u}^{+})$ denotes a numerically
calculated boundary flux, depending on the element-interior velocity $\bm{u}^-$
and its neighbour's velocity $\bm{u}^+$. The choice of such a flux is
solver-specific and may invovle an upwinding approach or use of an appropriate
Riemann solver. Where second-order diffusive terms are required, \nek supports
the use of a local discontinous Galerkin (LDG) approach to minimize the stencil
required for communication~(see \cite{cockburn1998local} and
\cite[\S7.5.2]{KaSh05}). From a solver perspective, the implementation of the
above is fairly generic, requiring only the evaluation of the flux term
$\tilde{\bm{f}}$, conservation law $\bm{F}(\bm{u})$ and right-hand side source
terms $\bm{G}(\bm{u})$.

\subsection{Recap of \nek implementation}

In this section, we briefly outline the implementation of these methods inside
\nek. Further details on the overall design of \nek, as well as examples of how
to use it, can be found in the previous publication~\cite{nektarpp2015}.

The core of \nek comprises six libraries which are designed to emulate the
general mathematical formulation expressed above. They describe the problem in a
hierarchical manner, by working from elemental basis functions and shapes
through to a C++ representation of a high-order field and complete systems of
partial differential equations on a complex computational domain. Depending on
the specific application, this then allows developers to choose an appropriate
level for their needs, balancing ease of use at the highest level with
fine-level implementation details at the lowest. A summary of each library's
purpose is the following:
\begin{itemize}
  \item \texttt{LibUtilites}: elemental basis functions $\psi_p$, quadrature
  point distributions $\xi_i$ and basic building blocks such as I/O handling;
  \item \texttt{StdRegions}: reference regions $\mathcal{E}$ along with the
  definition of key finite element operations across them, such as integration,
  differentiation and transformations;
  \item \texttt{SpatialDomains}: the geometric mappings $\chi_e$ and factors
  $\frac{\partial\chi}{\partial\xi}$, as well as Jacobians of the mappings and
  the construction of the topology of the mesh from the input geometry;
  \item \texttt{LocalRegions}: physical regions in the domain, composing a
  reference region $\mathcal{E}$ with a map $\chi_e$, extensions of core
  operations onto these regions;
  \item \texttt{MultiRegions}: list of physical regions comprising $\Omega$,
  global assembly maps which may optionally enforce continuity, construction and
  solution of global linear systems, extension of local core operations to
  domain-level operations; and
  \item \texttt{SolverUtils}: building blocks for developing complete solvers.
\end{itemize}
In version 5.0, four additional libraries have been included. Each of these can
be seen as a non-core, in the sense that they provide additional functionality
to the core libraries above:
\begin{itemize}
  \item \texttt{Collections}: encapsulates the implementation of key kernels
  (such as inner products and transforms) with an emphasis on evaluating
  operators collectively for similar elements;
  \item \texttt{GlobalMapping}: implements a mapping technique that allows
  quasi-3D simulations (i.e. those using a hybrid 2D spectral element/1D Fourier
  spectral discretisation) to define spatially-inhomogeneous deformations;
  \item \texttt{NekMeshUtils}: contains interfaces for CAD engines and key mesh
  generation routines, to be used by the \emph{NekMesh} mesh generator; and
  \item \texttt{FieldUtils}: defines various post-processing modules that can be
  used both by the post-processing utility \emph{FieldConvert}, as well as
  solvers for in-situ processing.
\end{itemize}
We describe the purpose of these libraries in greater detail in
Sections~\ref{sec:collections}, \ref{sec:mapping}, \ref{sec:nekmesh}
and~\ref{sec:insitu} respectively.


\section{Software and Performance Developments}
\label{sec:software}

This section reviews the software and performance developments added to \nek
since our last major release. We note that a significant change from previous
releases is the use of \emph{C++11}-specific language features throughout the
framework. A brief summary of our changes in this area include:

\begin{itemize}
  \item transitioning from various data structures offered in {\tt boost} to
  those now natively available in \emph{C++11}: in particular, smart pointers
  such as {\tt shared\_ptr}, unordered STL containers and function bindings;
  \item avoid the use of {\tt typedef} aliases for complex data structure types,
  in deference to the use of {\tt auto} where appropriate;
  \item similarly, where appropriate we make use of range-based {\tt for} loops
  to avoid iterator {\tt typedef} usage and simplify syntax;
  \item use of variadic templates in core memory management and data structures
  to avoid the use of syntax-dense preprocessing macros at compile time.
\end{itemize}

Alongside the many developments outlined here, the major change in the \nek API
resulting from this switch has further motivated the release of a new major
version of the code.
The developments described in this section are primarily geared towards our
continuing efforts to support our users on large-scale high-performance
computing (HPC) systems.

\subsection{Parallel I/O}
\label{sec:parallel_io}

Although the core of {\nek} has offered efficient parallel scaling for some time (as reported in previous
work~\cite{nektarpp2015}), one aspect that has been improved substantially in
the latest release is support for parallel I/O, both during the setup phase of
the simulation and when writing checkpoints of field data for unsteady
simulations. In both cases, we have added support for new, parallel-friendly
mesh input files and data checkpoint files that use the HDF5 file
format~\cite{folk2011overview}, in combination with Message Passing
Interface (MPI) I/O, to significantly
reduce bottlenecks relating to the use of parallel filesystems. This approach enables \nek to either read or
write a single file across all processes, as opposed to a file-per-rank output
scheme that can place significant pressure on parallel filesystems, particularly during the
partitioning phase of the simulation. Here we discuss the
implementation of the mesh input file format; details regarding the field output
can be found in~\cite{barefordimproving}.

One of the key challenges identified in the use of {\nek} within large-scale HPC
environments is the use of an XML-based input format used for defining the mesh
topology. Although XML is highly convenient from the perspective of readability
and portability, particularly for small simulations, the format does pose a
significant challenge at larger scales since, when running in parallel, there is
no straightforward way to extract a part of an XML file on each process. This
means that in the initial phase of the simulation, where the mesh is partitioned
into smaller chunks that run on each process, there is a need for at least one
process to read the entire XML file. Even at higher orders, where meshes are
typically coarse to reflect the additional degrees of freedom in each
element, detailed simulations of complex geometries typically require large,
unstructured meshes of millions of high-order elements. Having only a single process read
this file therefore imposes a natural limit to the strong scaling of the setup
phase of simulations -- that is, the maximum number of processes that can be
used -- due to the large memory requirement and processing time to produce
partitioned meshes. It also imposes potentially severe restrictions on start-up
time of simulations and/or the post-processing of data, hindering throughput for
very large cases.

\begin{figure*}
  \begin{subfigure}[b]{0.48\textwidth}
    \begin{center}
      \begin{tikzpicture}
        \draw[black,thick] (0,0) rectangle (4,4);
        \fill[black] (0,0) circle (3pt) node[below] {\color{red} vertex 0};
        \fill[black] (4,0) circle (3pt) node[below] {\color{red} vertex 1};
        \fill[black] (4,4) circle (3pt) node[above] {\color{red} vertex 2};
        \fill[black] (0,4) circle (3pt) node[above] {\color{red} vertex 3};
        \node[above] at (2,0) {\color{blue}edge 0};
        \node[right] at (4,2) {\color{blue}edge 1};
        \node[below] at (2,4) {\color{blue}edge 2};
        \node[left] at  (0,2) {\color{blue}edge 3};
      \end{tikzpicture}
    \end{center}
    \caption{A simple two-dimensional quadrilateral mesh consisting of a single
      element.}
  \end{subfigure}
  \begin{subfigure}[b]{0.48\textwidth}%
    \begin{center}
      \includegraphics{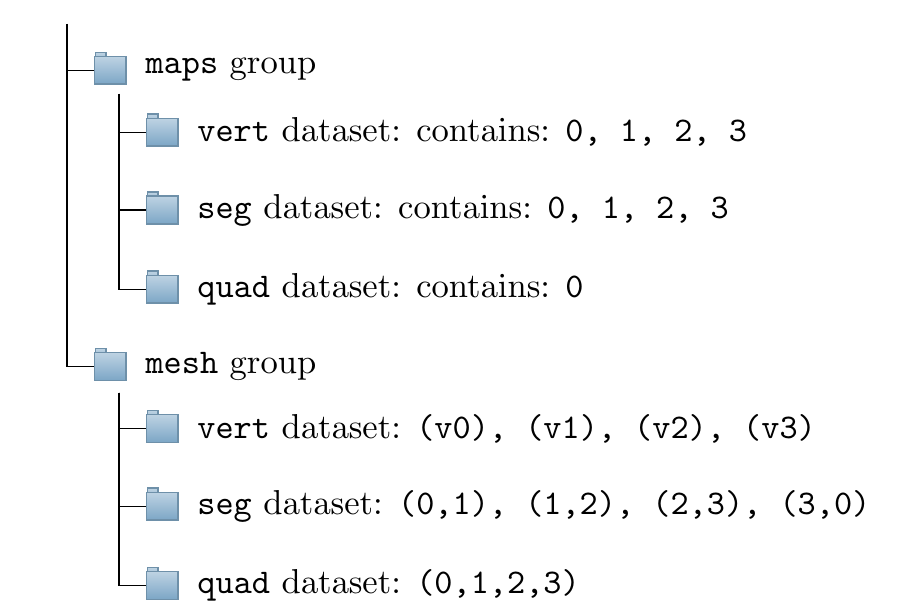}
    \end{center}
    \caption{Directory-dataset structure that is used for storage of the
      topological data in the left-hand figure.}
  \end{subfigure}
  
  \caption{An example of a single quadrilateral element grid. In (a), we show
    the topological decomposition of the element into its 4 edges and
    vertices. Figure (b) shows a schematic of the filesystem-type structure
    implemented in HDF5 that is used for storage of this topological
    information.}
  \label{fig:hdf5-quad}
\end{figure*}

Although various approaches have been used to partially mitigate this
restriction, such as pre-partitioning the mesh before the start of the
simulation and utilising a compressed XML format that reduces file sizes and the
XML tree size, these do not themselves cure the problem entirely. In the latest
version of {\nek} we address this issue with a new Hierarchical Data Format v5 (HDF5) file format. To
design the layout of this file, we recall that the structure of a basic {\nek} mesh requires
the following storage:
\begin{itemize}
  \item Vertices of the mesh are defined using double-precision floating point
  numbers for their three coordinates. Each vertex has a unique ID.
  \item All other elements are defined using integer IDs in a hierarchical
  manner; for example in three dimensions edges are formed from pairs of
  vertices, faces from $3$ or $4$ edges and elements from a collection of faces.
\end{itemize}
This hierarchical definition clearly aligns with the HDF5 data layout. To
accommodate the `mapping' of a given ID into a tuple of IDs or vertex
coordinates, we adopt the following basic structure:
\begin{itemize}
  \item The {\tt mesh} group contains multi-dimensional datasets that define the
  elements of a given dimension. For example, given $N$ quadrilaterals, the {\tt
    quad} dataset within the {\tt mesh} group is a $N \times 4$ dataset of
  integers, where each row denotes the 4 integer IDs of edges that define that
  quadrilateral.
  \item The {\tt maps} group contains one-dimensional datasets that define the
  IDs of each row of the corresponding two-dimensional dataset inside {\tt
    mesh}.
\end{itemize}
An example of this structure for a simple quadrilateral mesh is given in
Figure~\ref{fig:hdf5-quad}. We also define additional datasets to define element
curvature and other ancillary structures such as boundary regions.

When running in parallel, \nek adopts a domain decomposition strategy, whereby
the mesh is partitioned into a subset of the whole domain for each process. This
can be done either at the start of the simulation, or prior to running it.
Parallelisation is achieved using the standard MPI protocol, where each process
is independently executed and there is no explicit use of shared memory in
program code. Under the new HDF5 format, we perform a parallel partitioning
strategy at startup, which runs as follows:

\begin{itemize}
  \item Each process is initially assigned a roughly equal number of elements to
  read. This is calculated by querying the size of each elemental dataset to
  determine the total number of elements, and then partitioned equally according
  to the rank of the process and total number of processors.
  \item The dual graph corresponding to each process' subdomain is then
  constructed. Links to other process subdomains are established by using
  ghost nodes to those process' nodes.
  \item The dual graph is passed to the PT-Scotch library~\cite{chevalier-2008}
  to perform partitioning in parallel on either the full system or a subset of
  processes, depending on the size of the graph.
  \item Once the resulting graph is partitioned, the datasets are read in
  parallel using a top-down process: i.e. in three dimensions, we read the
  volumes, followed by faces, edges and finally vertices. In the context of
  Figure~\ref{fig:hdf5-quad}, this would consist of reading the {\tt quad}
  dataset, followed by the {\tt seg} dataset, followed by the {\tt vert}
  dataset.
  \item Note that at each stage, each processor only reads the geometric
  entities that are required for its own partition, which is achieved through
  the use of HDF5 selection routines when reading the datasets.
  \item The \nek geometry objects are then constructed from these data in a
  bottom-up manner: i.e. vertices, followed by edges, followed by faces and
  finally volumes, as required by each processor.
  \item This concludes the construction of the linear mesh: curvature
  information is stored in separate datasets, and is also read at this stage as
  required for each element.
  \item Finally, ancillary information such as composites and domain definition
  are read from the remaining datasets.
\end{itemize}

The new HDF5 based format is typically significantly faster than the existing
XML format to perform the initial partitioning phase of the simulation. Notably,
whereas execution times for the XML format increase with the number of nodes
being used (likely owing to the file that must be written for each rank by the
root processor), the HDF5 input time remains roughly constant. We note that the
HDF5 format also provides benefits for the post-processing of large simulation
data, as the {\em FieldConvert} utility is capable of using this format for
parallel post-processing of data.

\subsection{In-situ processing}
\label{sec:insitu}
The increasing capabilities of high-performance computing facilities allow us to
perform simulations with a large number of degrees of freedom which leads to
challenges in terms of post-processing. The first problem arises when we
consider the storage requirements of the complete solution of these
simulations. Tasks such as generating animations, where we need to consider the
solution at many time instances, may become infeasible if we have to store the
complete fields at each time instance. Another difficulty occurs due to the
memory requirements of loading the solution for post-processing. Although this
can be alleviated by techniques such as subdividing the data and processing one
subdivision at a time, this is not possible for some operations requiring global
information, such as performing a $C^0$-projection that involves the inversion
of a global mass matrix. In such cases, the memory requirements might force the
user to perform post-processing using a number of processing nodes similar to
that used for the simulation.

To aid in dealing with this issue, \nek now supports processing the solution
{\em in situ} during the simulation. The implementation of this feature was
facilitated by the modular structure of our post-processing tool, {\em
  FieldConvert}. This tool uses a pipeline of modules, passing mesh and field
data between them, to arrive at a final output file. This comprises one or more
input modules (to read mesh and field data), zero or more processing modules (to
manipulate the data, such as calculating derived quantities or extracting
boundary information), and a single output module (to write the data in one of a
number of field and visualisation formats). To achieve {\em in situ} processing,
{\em FieldConvert} modules were moved to a new library ({\tt FieldUtils}),
allowing them to be executed during the simulation as well as shared with the
{\em FieldConvert} utility. The actual execution of the modules during \emph{in
  situ} processing is performed by a new subclass of the {\tt Filter} class,
which is called periodically after a prescribed number of time-steps to perform
operations which do not modify the solution field. This filter structure allows
the user to choose which modules should be used and to set configuration
parameters. Multiple instances of the filter can be used if more than one
post-processing pipeline is desired.

There are many example applications for this new feature. The most obvious is to
generate a field or derived quantity, such as vorticity, as the simulation is
running. An example of this is given in the supplementary materials
Example~\ref{f:suppl:cylinder}, in which the vorticity is calculated every 100
timesteps whilst removing the velocity and pressure fields to save output file
space, using the following {\tt FILTER} configuration in the session file:
\begin{lstlisting}[style=XMLStyle,gobble=2]
  <FILTER TYPE="FieldConvert">
    <PARAM NAME="OutputFile"> vorticity.vtu </PARAM>
    <PARAM NAME="OutputFrequency"> 100 </PARAM>
    <PARAM NAME="Modules">
      vorticity
      removefield:fieldname=u,v,p
    </PARAM>
  </FILTER>
\end{lstlisting}
This yields a number of parallel-format block-unstructured VTK files (the VTU
format), as described in~\cite{schroeder2004visualization}, that can be
visualised in appropriate such as Paraview~\cite{ahrens2005paraview} and
subsequently assembled to form an animation. Other example applications include
extracting slices or isocontours of the solution at several time instants for
creating an animation. Since the resulting files are much smaller than the
complete solution, there are significant savings in terms of storage when
compared to the traditional approach of obtaining checkpoints which are later
post-processed. Another possibility is to perform the post-processing operations
after the last time-step, but before the solver returns. This way, it is
possible to avoid the necessity of starting a new process which will have to
load the solution again, leading to savings in computing costs.

\subsection{Collective linear algebra operations}
\label{sec:collections}
One of the primary motivations for the use of high-order methods is their
ability to outperform standard linear methods on modern computational
architectures in terms of equivalent error per degree of freedom. Although the
cost in terms of floating point operations (FLOPS) of calculating these degrees
of freedom increases with polynomial order, the dense, locally-compact structure
of higher-order operators lends itself to the current hardware environment, in
which FLOPS are readily available but memory bandwidth is highly limited. In
this setting, the determining factor in computational efficiency, or ability to
reach peak performance of hardware, is the arithmetic intensity of the method;
that is, the number of FLOPS performed for each byte of data transferred over
the memory bus. Algorithms need to have high arithmetic
intensity in order to fully utilise the computing power of modern computational
hardware.

However, the increase in FLOPS at higher polynomial orders must be balanced
against the desired accuracy so that execution times are not excessively
high. An observation made early in the development of spectral element methods
is that operator counts can be substantially reduced by using a combination of a
tensor product basis, together with a tensor contraction technique referred to
as \emph{sum-factorisation}. This technique, exploited inside of \nek as well as
other higher-order frameworks such as deal.II~\cite{BaHaKa07}, uses a small
amount of temporary storage to reduce operator counts from $\mathcal{O}(P^{2d})$
to $\mathcal{O}(P^{d+1})$ at a given order $P$. For example, consider a
polynomial interpolation on a quadrilateral across a tensor product of
quadrature points $\xi = (\xi_{1i}, \xi_{2j})$, where the basis admits a tensor
product decomposition $\phi_{pq}(\xi) = \phi_p(\xi_1)\phi_q(\xi_2)$. This
expansion takes the form
\begin{align*}
  u^{\delta}(\xi_{1i},\xi_{2j}) &= \sum_{p=0}^P\sum_{q=0}^Q \hat{u}_{pq}
  \phi_p(\xi_{1i})\phi_q(\xi_{2j}) \\
  &= \sum_{p=0}^P\phi_p(\xi_{1i}) \left[\sum_{q=0}^Q \hat{u}_{pq}
  \phi_q(\xi_{2j})\right].
\end{align*}
By precomputing the bracketed term and storing it for each $p$ and $j$, we can
reduce the number of floating point operations from $\mathcal{O}(P^4)$ to
$\mathcal{O}(P^3)$. One of the distinguishing features of \nek is that these
types of basis functions are defined not only for tensor-product quadrilaterals
and hexahedra, but also unstructured elements (triangles, tetrahedra, prisms and
pyramids) through the use of a collapsed coordinate system and appropriate basis
functions. For more details on this formulation, see~\cite{KaSh05}.

The efficient implementation of the above techniques on computational hardware
still poses a significant challenge for practitioners of higher-order methods. For example,
\nek was originally designed using a hierarchical, inheritance-based approach,
where memory associated with elemental degrees of freedom is potentially
scattered non-contiguously in memory. Although this was more appropriate at the
initial time of development a decade ago, in modern terms this does not align
with the requirements for optimal performance, in which large blocks of
memory should be transferred and as many operations acted on sequentially across
elements, so as to reduce memory access and increase data locality and cache
usage. The current efforts of the development team are therefore focused on
redesigns to the library to accommodate this process. In particular, since
version 4.1, \nek has included a library called {\tt Collections} which is
designed to provide this optimisation. In the hierarchy of Nektar++ libraries, {\tt Collections} sits
between {\tt LocalRegions}, which represent individual elements, and {\tt
  MultiRegions}, which represent their connection in either a $C^0$ or
discontinuous Galerkin setting. The purpose of the library, which is described
fully in~\cite{moxey-2016b}, is to facilitate optimal linear algebra strategies
for large groupings of elements that are of the same shape and utilise the same
basis. To facilitate efficient execution across a
broad range of polynomial orders, we then consider a number of implementation
strategies including:
\begin{itemize}
  \item {\bf StdMat}: where a full-rank matrix of the operator on a standard
  element is constructed, so that the operator can be evaluated with a single
  matrix-matrix multiplication;
  \item {\bf IterPerExp}: where the sum-factorisation technique is evaluated
  using an iteration over each element, but geometric factors (e.g.
  $\partial \mathbf{x}/\partial \xi$) are amalgamated between elements; and
  \item {\bf SumFac}: where the sum-factorisation technique is evaluated
  across multiple elements concurrently.
\end{itemize}
This is then combined with an autotuning strategy, run at simulation startup,
which attempts to identify the fastest evaluation strategy depending on characteristics of the computational mesh and basis parameters. Autotuning can be enabled in any simulation through the
definition of an appropriate tag inside the {\tt NEKTAR} block that defines a
session file:
\begin{lstlisting}[style=XMLStyle,gobble=2]
  <COLLECTIONS DEFAULT="auto" />
\end{lstlisting}
A finer-grained level of control over the Collections setup and implementation
strategies is documented in the user guide. Performance improvements using collections
are most readily seen in fully-explicit codes such as the {\tt
  CompressibleFlowSolver} and {\tt AcousticSolver}. The vortex pair example
defined in Section~\ref{sec:acoustic} and provided in
Example~\ref{f:suppl:user-guide} demonstrates the use of the collections
library.

\subsection{Solver coupling}
\label{sec:coupling}
The \nek framework was extended with a coupling interface \cite{Lackhove2018}
that enables sending and receiving arbitrary variable fields at run time.  Using
such a technique, a coupling-enabled solver can exchange data with
other applications to model multi-physics problems in a co-simulation setup.
Currently, two coupling interfaces are available within \nek; a file-based
system for testing purposes, and an MPI-based implementation for large-scale HPC
implementations.  The latter was designed to facilitate coupling \nek solvers
with different software packages which use other
discretization methods and operate on vastly different time- and length-scales.
To couple two incompatible discretisations, an intermediary expansion is used
which can serve as a projection between both sides of the field.  Coupling is
achieved by introducing an intermediate expansion, which uses the same
polynomial order and basis definitions as the parent \nek{} solver; however, a
continuous projection and a larger number of quadrature points than the original
expansion of the \nek{} solver are used.  Based on this intermediate
representation, the coupling strategy is comprised of three major steps:
\begin{itemize}
  \item {\bf Step 1:} The field values are requested from the sending
  application at the intermediate expansion's quadrature points.  Here, aliasing
  can be effectively avoided by an appropriate selection of quadrature order and
  distribution.  Point values that lie outside of the senders' computational
  domain can be either replaced by a default value or extrapolated from their
  available nearest neighbour.

  \item {\bf Step 2:} The physical values at the quadrature points are then
  transformed into modal space.  This is achieved by a modified forward
  transform that involves the differential low-pass filter \cite{Germano1986a}:
  \begin{equation} \label{eq:filtereq}
    u^{**}  - \left(\frac{\Delta \lambda}{2 \pi}\right)^2 \nabla^2 u^{**} = u^*
    \, , \quad \left.\frac{\partial u^{**} }{\partial x_i } \right|_{\partial\Omega } = 0 
  \end{equation}
  where $u^{*}$ denotes the received field, $u^{**}$ the filtered field and
  $\Delta \lambda$ the user specified filter width.  The filter removes small
  scale features {\em a priori} and thus reduces the error associated with the
  transform.  Moreover, it does not add unwanted discontinuities at the element
  boundaries and imposes a global smoothing, due to the continuity of the
  intermediate expansion.

  \item {\bf Step 3:} A linear interpolation in time can be performed to
  overcome larger time scales of the sending application.  Due to their
  identical expansion bases and orders, the resulting coefficients can be
  directly used in the original expansion of the solver.
\end{itemize}

As is evident from the above strategy, sending fields to other solvers only
requires an application to provide discrete values at the requested
locations. In \nek, this can be achieved by evaluating the expansions or by a
simpler approximation from the immediately available quadrature point values.
All processing is performed by the receiver. The complex handling of data
transfers is accomplished by the open-source CWIPI library~\cite{Refloch2011},
which enables coupling of multiple applications by using decentralized
communication.  It is based purely on MPI calls, has bindings for C, Fortran and
Python, handles detection of out-of-domain points and has been shown to exhibit
good performance~\cite{Duchaine2015}.  With only CWIPI as a dependency and a
receiver-centric strategy that can be adjusted to any numerical setup, the
implementation of compatible coupling interfaces is relatively straightforward.

\begin{figure*}[htb]
    \centering
    \footnotesize
    \def\svgwidth{\linewidth}
    \begingroup%
    \makeatletter%
    \providecommand\color[2][]{%
      \errmessage{(Inkscape) Color is used for the text in Inkscape, but the package 'color.sty' is not loaded}%
      \renewcommand\color[2][]{}%
    }%
    \providecommand\transparent[1]{%
      \errmessage{(Inkscape) Transparency is used (non-zero) for the text in Inkscape, but the package 'transparent.sty' is not loaded}%
      \renewcommand\transparent[1]{}%
    }%
    \providecommand\rotatebox[2]{#2}%
    \newcommand*\fsize{\dimexpr\f@size pt\relax}%
    \newcommand*\lineheight[1]{\fontsize{\fsize}{#1\fsize}\selectfont}%
    \ifx\svgwidth\undefined%
    \setlength{\unitlength}{476.13059217bp}%
    \ifx\svgscale\undefined%
    \relax%
    \else%
    \setlength{\unitlength}{\unitlength * \real{\svgscale}}%
    \fi%
    \else%
    \setlength{\unitlength}{\svgwidth}%
    \fi%
    \global\let\svgwidth\undefined%
    \global\let\svgscale\undefined%
    \makeatother%
    \begin{picture}(1,0.1748752)%
      \lineheight{1}%
      \setlength\tabcolsep{0pt}%
      \put(0,0){\includegraphics[width=\unitlength,page=1]{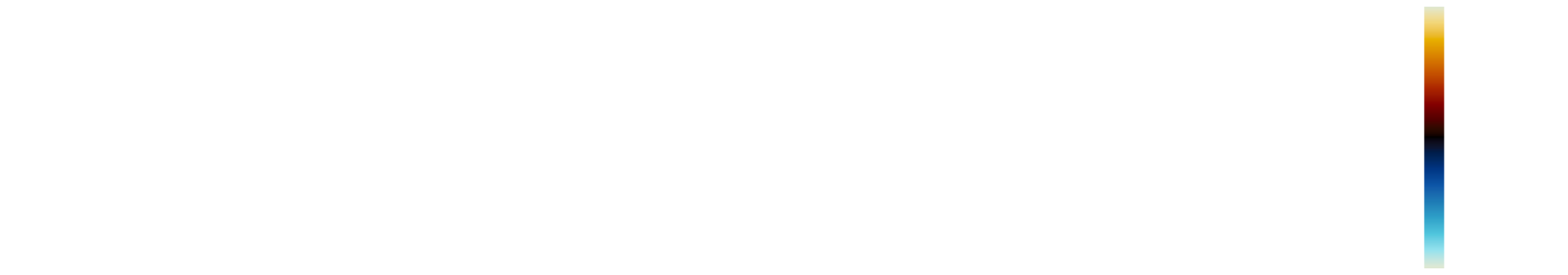}}%
      \put(0.98018817,0.08740563){\color[rgb]{0,0,0}\rotatebox{-90}{\makebox(0,0)[t]{\lineheight{1.25000012}\smash{\begin{tabular}[t]{c}$\dot{\omega}_\mathrm{c}$ [Pa/s]\end{tabular}}}}}%
      \put(0,0){\includegraphics[width=\unitlength,page=2]{F_0_p_CAA_CFD.pdf}}%
      \put(0.93184702,0.16648898){\color[rgb]{0,0,0}\makebox(0,0)[lt]{\lineheight{1.25}\smash{\begin{tabular}[t]{l}1.5E8\end{tabular}}}}%
      \put(0.93202636,0.11101102){\color[rgb]{0,0,0}\makebox(0,0)[lt]{\lineheight{1.25}\smash{\begin{tabular}[t]{l}5E7\end{tabular}}}}%
      \put(0.931647,0.08339535){\color[rgb]{0,0,0}\makebox(0,0)[lt]{\lineheight{1.25}\smash{\begin{tabular}[t]{l}0\end{tabular}}}}%
      \put(0.93172893,0.0556629){\color[rgb]{0,0,0}\makebox(0,0)[lt]{\lineheight{1.25}\smash{\begin{tabular}[t]{l}-5E7\end{tabular}}}}%
      \put(0.93120596,0.00016){\color[rgb]{0,0,0}\makebox(0,0)[lt]{\lineheight{1.25}\smash{\begin{tabular}[t]{l}-1.5E8\end{tabular}}}}%
      \put(0,0){\includegraphics[width=\unitlength,page=3]{F_0_p_CAA_CFD.pdf}}%
      \put(0.01672286,0.01384768){\color[rgb]{1,1,1}\makebox(0,0)[t]{\lineheight{1.25}\smash{\begin{tabular}[t]{c}$\Delta \lambda$\end{tabular}}}}%
      \put(0,0){\includegraphics[width=\unitlength,page=4]{F_0_p_CAA_CFD.pdf}}%
    \end{picture}%
    \endgroup%
    \caption{Instantaneous acoustic source term as represented in CFD
      (proprietary finite volume flow solver with $\Delta h < 1.4\text{mm}$
      mesh) and CAA (\nek{} AcousticSolver with $\Delta h = 20\text{mm}$ mesh
      and fourth order expansion). Slice through a three-dimensional domain.}
    \label{fig:thermosources}
\end{figure*}

An example result of a transferred field is given in
Figure~\ref{fig:thermosources}.  For a hybrid noise
simulation~\cite{Lackhove2018}, the acoustic source term depicted at the top was
computed by a proprietary, finite volume flow solver on a high-resolution mesh
($\Delta h < 1.4\text{mm}$) and transferred to the \nek {\tt AcousticSolver},
which we describe in Section~\ref{sec:acoustic}.  After sampling, receiving,
filtering, projection and temporal interpolation, the extrema of the source term
are cancelled out and blurred by the spatial filter.  Consequently, a much
coarser mesh ($\Delta h = 20\text{mm}$) with a fourth order expansion is
sufficient for the correct representation of the resulting field, which
significantly reduces the computational cost of the simulation.  The
corresponding loss of information is well defined by the filter width
$\Delta \lambda$ and limited to the high-frequency range, which is irrelevant
for the given application.


\subsection{Python interface}
\label{sec:python}
Although \nek is designed to provide a modern {\em C++} interface to high-order
methods, its use of complex hierarchies of classes and inheritance, as well as
the fairly complex syntax of {\em C++} itself, can lead to a significant barrier
to entry for new users of the code. At the same time, the use of Python in
general scientific computing applications, and data science application areas in
particular, is continuing to grow, in part due to its relatively simple syntax
and ease of use. Additionally, the wider Python community offers a multitude of
packages and modules to end users, making it an ideal language through which
disparate software can be `glued' to perform very complex tasks with relative
ease. For the purposes of scientific computing codes, the Python {\em C} API
also enables the use of higher-performance compiled code, making it suitable in
instances where interpreted pure Python would be inefficient and impractical, as
can be seen using packages such as {\tt NumPy} and {\tt SciPy}. These factors
therefore make Python an ideal language through which to both introduce new
users to a complex piece of software, interact with other software packages and,
at the same time, retain a certain degree of performance that would not be
possible from a purely interpreted perspective.

The Version 5.0 release of {\nek} offers a set of high-level bindings for a
number of classes within the core {\nek} libraries. The purpose of these
bindings is to significantly simplify the interfaces to key {\nek} libraries,
offering both a teaching aide for new users to the code, as well as a way to
connect with other software packages and expand the scope of the overall
software. To achieve this, we leverage the Boost.Python
package~\cite{abrahams2003building}, which offers a route to handling many of
the complexities and subtleties of translating {\em C++} functions and classes
across to the Python {\em C} API. A perceived drawback of this approach is the
lack of automation. As Boost.Python is essentially a wrapper around the Python
{\em C} API, any bindings must be handwritten, whereas other software such as
f2py~\cite{peterson2009f2py} or SWIG~\cite{beazley1996swig} offer the ability to
automatically generate bindings from the {\em C++} source. However, our
experience of this process has been that, other than implementation effort,
handwritten wrappers provide higher quality and more stability, particularly
when combined with an automated continuous integration process as is adopted in
{\nek}, as well as better interoperability with key Python packages such as {\tt
  numpy}. In our particular case, heavy use of {\em C++11} features such as {\tt
  shared\_ptr} and the {\nek} {\tt Array} class for shared storage meant that
many automated solutions would not be well-suited to this particular
application.

An example of the Python bindings can be seen in Listing~\ref{lst:python}, where
we perform the Galerkin projection of the smooth function
$f(x,y) = \cos(x)\cos(y)$ onto a standard quadrilateral expansion at order $P=7$,
using $P+1$ Gauss-Lobatto-Legendre quadrature points to exactly integrate the
mass matrix. We additionally perform an integral of this function (whose exact
value is $4\sin^2(1)$).  As can be seen in this example, the aim of the bindings
is to closely mimic the layout and structure of the {\em C++} interface, so that they
can be used as a learning aide for to the full {\em C++} API. Additionally, the Python
bindings make full use of Boost.Python's automatic datatype conversion
facilities. In particular, significant effort has been extended to facilitate
seamless interaction between the {\tt NumPy.ndarray} class, which is almost
universally used in Python scientific computing applications for data storage,
and the \nek storage {\tt Array<OneD,~*>} classes. This allows an {\tt ndarray}
to be passed into {\nek} functions directly and vice versa. Moreover this
interaction uses the Boost.Python interface to {\tt NumPy} to ensure that
instead of copying data (which could be rather inefficient for large arrays),
this interaction uses a shared memory space between the two data
structures. Reference counting is then used to ensure data persistence and
memory deallocation, depending on whether memory was first allocated within the
{\em C++} environment or Python.

\begin{python}[gobble=2,caption={Using the {\nek} 5.0 Python bindings to perform
    a simple Galerkin projection and integral on a standard quadrilateral element.},label={lst:python}]
  import NekPy.LibUtilities as LibUtil
  import NekPy.StdRegions as StdReg
  import numpy as np

  # Set P = 8 modes and Q = P + 1 quadrature points.
  nModes = 8
  nPts   = nModes + 1

  # Create GLL-distributed quadrature points.
  pType  = LibUtil.PointsType.GaussLobattoLegendre
  pKey   = LibUtil.PointsKey(nPts, pType)

  # Create modified C^0 basis on these points.
  bType  = LibUtil.BasisType.Modified_A
  bKey   = LibUtil.BasisKey(bType, nModes, pKey)

  # Create quadrilateral expansion using this basis
  # in each coordinate direction (tensor product).
  quad   = StdReg.StdQuadExp(bKey, bKey)

  # L^2 projection of f(x,y) = cos(x)*cos(y) onto the
  # quadrilateral element. Note x,y are numpy ndarrays
  # and evaluation of cos() is performed using numpy.
  x, y   = quad.GetCoords()
  fx     = np.cos(x) * np.cos(y)
  proj   = quad.FwdTrans(fx)

  # Integrate function over the element.
  print("Integral = {:.4f}".format(quad.Integral(fx)))
\end{python}

\section{Developments in Numerical Methods}
\label{sec:numerics}

This section highlights our recent developments on numerical methods contained
with the \nek release.


\subsection{Method of moving frames}
\label{sec:mmf}
\begin{figure}[ht]
  \centering
  \subcaptionbox{Conservational laws}{\label{mmf1}\includegraphics[width = 3cm]{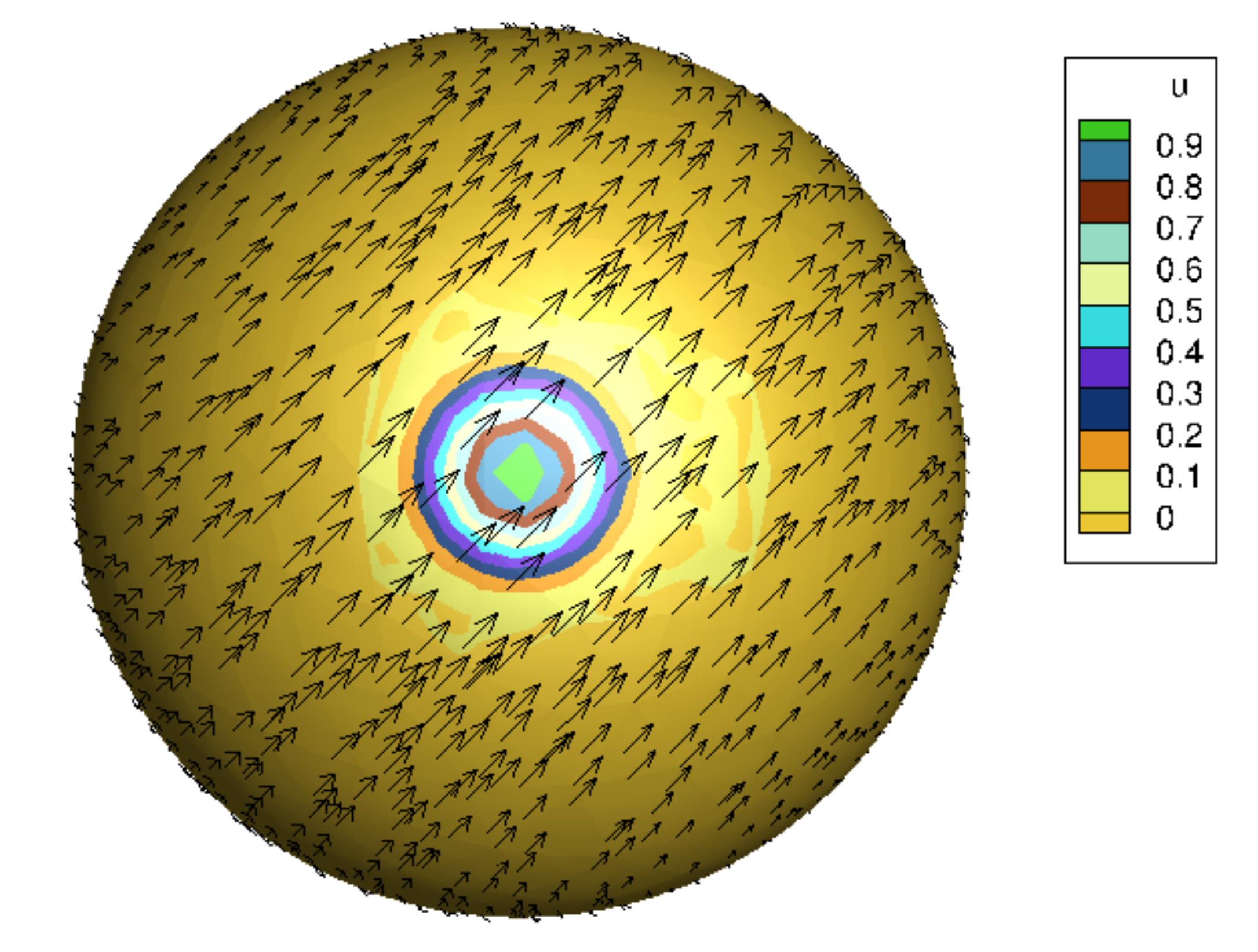}} \hspace{1cm}
  \subcaptionbox{Diffusion equation}{\label{mmf2}\includegraphics[width = 3cm]{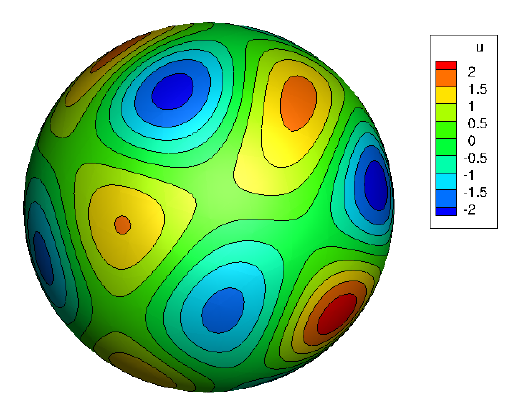}}
  \subcaptionbox{Shallow water equations}{\label{mmf3}\includegraphics[width = 3cm]{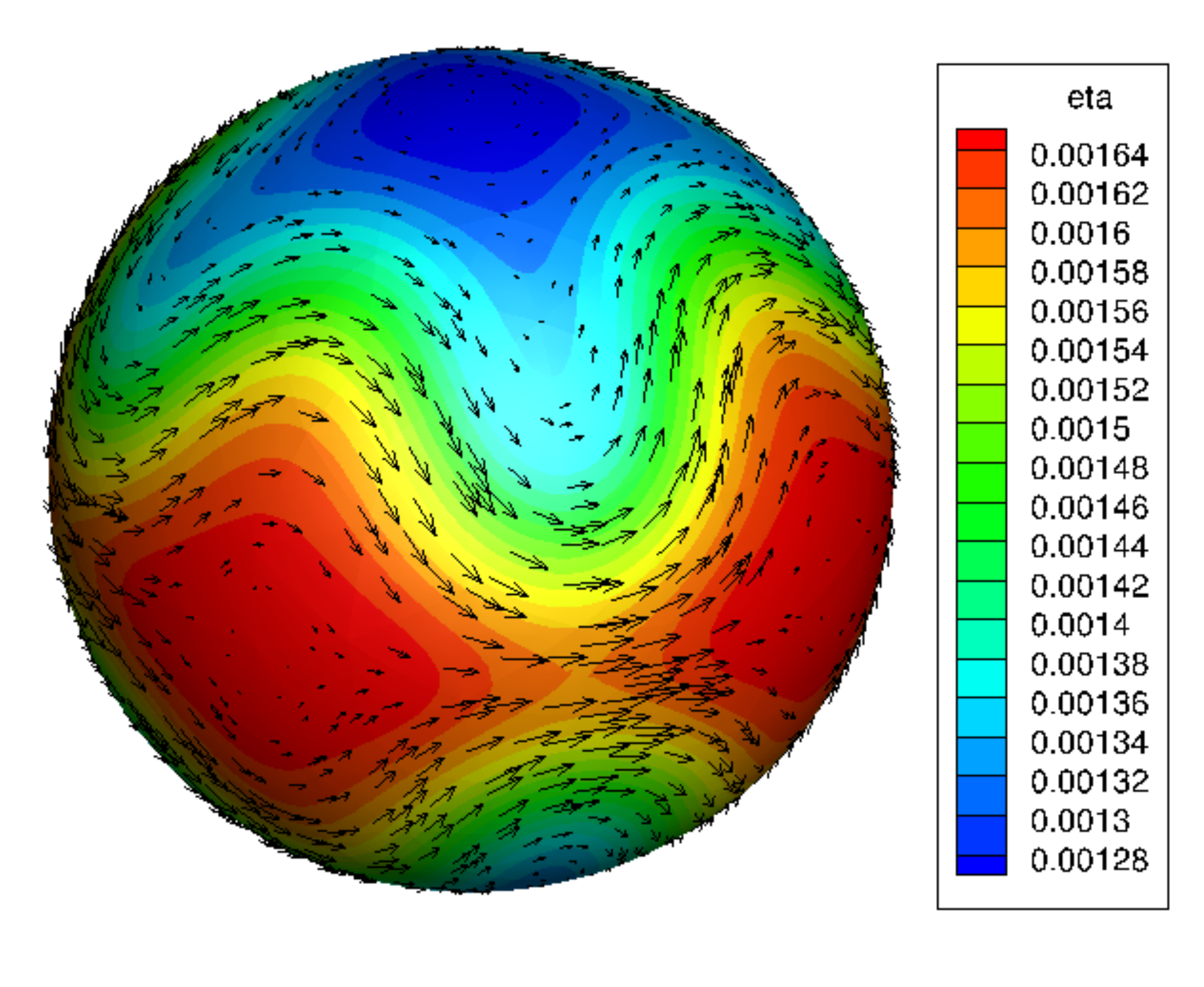}}  \hspace{1cm}
  \subcaptionbox{Maxwell's equations}{\label{mmf4}\includegraphics[width = 3cm]{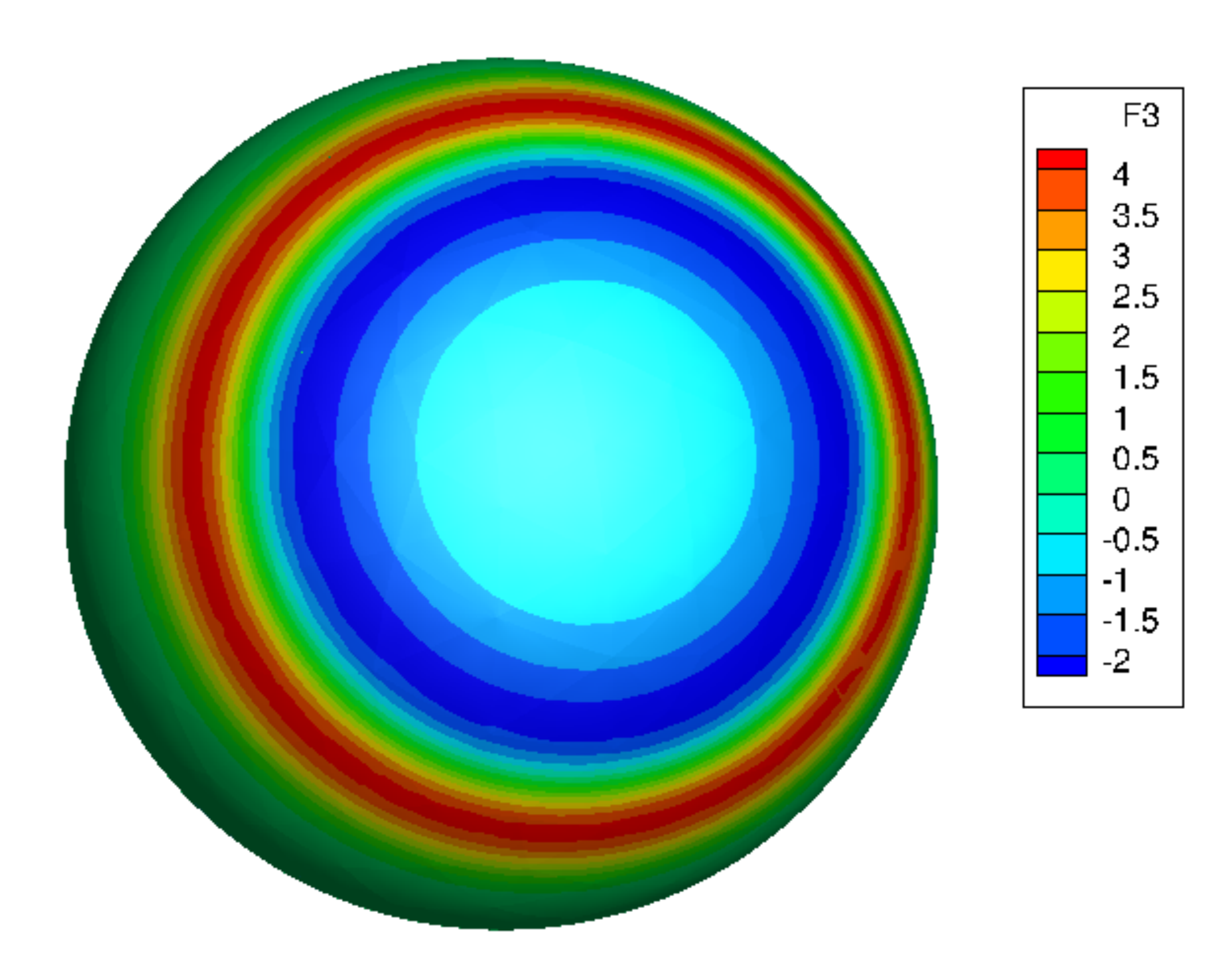}}
  \caption{Numerical simulation of the MMF scheme in {\nek} for several partial differential equations solved on the sphere.}
  \label {mmf}
\end{figure}

Modern scientific computation faces unprecedented demands on computational
simulation in multidimensional complex domains composed of various
materials. Examples of this include solving shallow water equations on a
rotating sphere for weather prediction, incorporating biological anisotropic
properties of cardiac and neural tissue for clinical diagnosis, and simulating
the electromagnetic wave propagation on metamaterials for controlling
electromagnetic nonlinear phenomena. All of these examples require the ability
to solve PDEs on manifolds embedded in higher-dimensional domains.  The method
of moving frames (MMF) implemented in \nek is a novel numerical scheme for
solving such computational simulations in geometrically-complex domains.

Moving frames, principally developed by \'{E}lie Cartan in the context of Lie
groups in the early 20th century~\cite{Cartan1,Cartan2,Cartan3}, are orthonormal
vector bases that are constructed at every grid point in the discrete space of a
domain $\Omega$.  Moving frames are considered as an `independent' coordinate
system at each grid point, and can be viewed as a dimensional reduction because
the number of moving frames corresponds to dimensionality, independent of the
space dimension. In this sense, `moving' does not mean that the frames are
time-dependent, but are different in a pointwise sense: i.e. if a particle
travels from one point to the other, then it may undergo a different frame,
which looks like a series of `moving' frames. More recently this approach has
been adapted more practical and computational purposes, mostly in computer
vision~\cite{Olver1998, Olver2001, Faugeras1994} and medical
sciences~\cite{Piuze2015}.

Building such moving frames is easily achieved by differentiating the
parametric mapping $\mathbf{x}$ of a domain element $\Omega_e$ with respect to
each coordinate axis of a standard reference space, followed by a Gram-Schmidt
orthogonalization process. We obtain orthonormal vector bases, denoted as
$\mathbf{e}^i$, with the following properties:
\begin{equation*}
\mathbf{e}^i \cdot \mathbf{e}^j = \delta^i_j , ~~~~~\| \mathbf{e}^i \| = 1,~~~~  1 \le i, j \le 3 ,
\end{equation*}
where $\delta^i_j$ denotes the Kronecker delta. Moreover, the moving frames are
constructed such that they are differentiable within each element and always lie on the tangent plane at each grid point. These two intrinsic
properties of frames implies that any vector or the gradient can be expanded on
moving frames as follows:
\begin{equation*}
  \mathbf{v} = v^1 \mathbf{e}^1 + v^2 \mathbf{e}^2,~~~~~~ \nabla u =  u^1 \mathbf{e}^1 + u^2 \mathbf{e}^2. 
\end{equation*}
Applying this expansion to a given PDE enables us to re-express it with moving
frames on any curved surface. Then, the weak formulation of the PDE with moving
frames, called the \textit{MMF scheme}, on a curved surface is similar to the
scheme in the Euclidean space, in the sense that it contains no metric tensor or
its derivatives and it does not require the construction of a continuous curved
axis in $\Omega$ which often produces geometric singularities. This is a direct
result of the fact that moving frames are \textit{locally} Euclidean. However,
the numerical scheme with moving frames results in the accurate solutions of
PDEs on any types of surfaces such as spheres, irregular surfaces, or non-convex
surfaces. Some examples of simulations that can be achieved under this approach
include conservational laws~\cite{MMF1}, the diffusion equation~\cite{MMF2},
shallow water equations (SWEs)~\cite{MMF3}, and Maxwell's
equations~\cite{MMF4}. Representative results from \nek for these equations on
the surface of a sphere are shown in Figure~\ref{mmf}.

Moreover, moving frames have been proven to be efficient for other geometrical
realizations, such as the representation of anisotropic properties of media on
complex domains~\cite{MMF2}, incorporating the rotational effect of any
arbitrary shape~\cite{MMF3}, and adapting isotropic upwind numerical flux in
anisotropic media~\cite{MMF4}. The accuracy of the MMF scheme with the
higher-order curvilinear meshes produced by {\em NekMesh}, described in
Section~\ref{sec:nekmesh}, is reported to be significantly improved for a high
$p$ and conservational properties such as mass and energy after a long time
integration, whereas the accuracy of the MMF-SWE scheme on {\em NekMesh} is
presented to be the best among all the previous SWE numerical
schemes~\cite{MMF5}. Ongoing research topics on moving frames are to construct
the connections of frames, to compute propagational curvature, and finally to
build an \textit{Atlas} (a geometric map with connection and curvature) in order
to provide a quantitative measurement and analysis of a flow on complex
geometry. Examples of ongoing research topics in this area are include
electrical activation in the heart~\cite{MMF6} and fiber tracking of white
matter in the brain.

\subsection{Spatially-variable polynomial order}
\label{sec:varp}
An important difficulty in the simulation of flows of practical interest is the
wide range of length- and time-scales involved, especially in the presence of
turbulence. This problem is aggravated by the fact that in many cases it is
difficult to predict where in the domain an increase in the spatial resolution
is required before conducting the simulation, while performing a uniform
refinement across the domain is computationally prohibitive. Therefore, in
dealing with these types of flows, it is advantageous to have an adaptive
approach which allows us to dynamically adjust the spatial resolution of the
discretisation both in time and in space.

Within the \shp element framework, it is possible to refine the solution
following two different routes. $h$-refinement consists of reducing the size of
the elements as would be done in low-order methods.  This is the common approach
for the initial distribution of degrees of freedom over the domain, with the
computational mesh clustering more elements in regions where small scales are
expected to occur, such as boundary layers. The other route is $p$-refinement
(sometimes called $p$-enrichment), where the spatial resolution is increased by
using a higher polynomial order to represent the solution. As discussed
in~\cite{nekpp-icosahom2016}, the polynomial order can be easily varied across
elements in the \shp element method if the expansion basis is chosen
appropriately. In particular if a basis admits a boundary-interior
decomposition, such as the modified $C^0$ basis described in
section~\ref{sec:methods} or the classical Lagrange interpolant basis, then the
variation in polynomial order can be built into the assembly operation between
interconnected elements. This allows for a simple approach to performing local
refinement of the solution, requiring only the adjustment of the polynomial
order in each element.

With this in mind, an adaptive polynomial order procedure has been implemented
in \nek, with successful applications to simulations of incompressible flows.
The basic idea in this approach is to adjust the polynomial order during the
solution based on an element-local error indicator.  The approach we used is
similar to that demonstrated for shock capturing in~\cite{PePe2006}, whereby
oscilliatory behaviour in the solution field is detected by an error sensor on
each element $\Omega_e$ calculated as
\[
  S_e = \frac{\|u_P - u_{P-1}\|^2_2}{\|u_P\|^2_2},
\]
where $u_P$ is the solution obtained for the $u$ velocity using the current
polynomial order $P$, $u_{P-1}$ is the projection of this solution to a
polynomial of order $P - 1$ and $\|\cdot\|_2$ denotes the $L^2$ norm. After each
$n_{\text{steps}}$ time-steps, this estimate is evaluated for each element. For
elements where the estimate of the error is above a chosen threshold, $P$ is
incremented by one, whereas in elements with low error $P$ is decremented by
one, respecting minimum and maximum values for $P$. The choice of
$n_{\text{steps}}$ is critical for the efficiency of this scheme, since it has
to be sufficiently large to compensate for the costs of performing the
refinement over a large number of time-steps, yet small enough to adjust to
changes in the flow. More details on this adaptive procedure for adjusting the
polynomial order, as well as its implementation in both CG and DG regimes, are
found in~\cite{nekpp-icosahom2016}.

\begin{figure*}
  \centering
  {%
    \setlength{\fboxsep}{0pt}%
    \subcaptionbox{Spanwise vorticity}[0.48\textwidth]{\fbox{\includegraphics[width=0.48\textwidth]{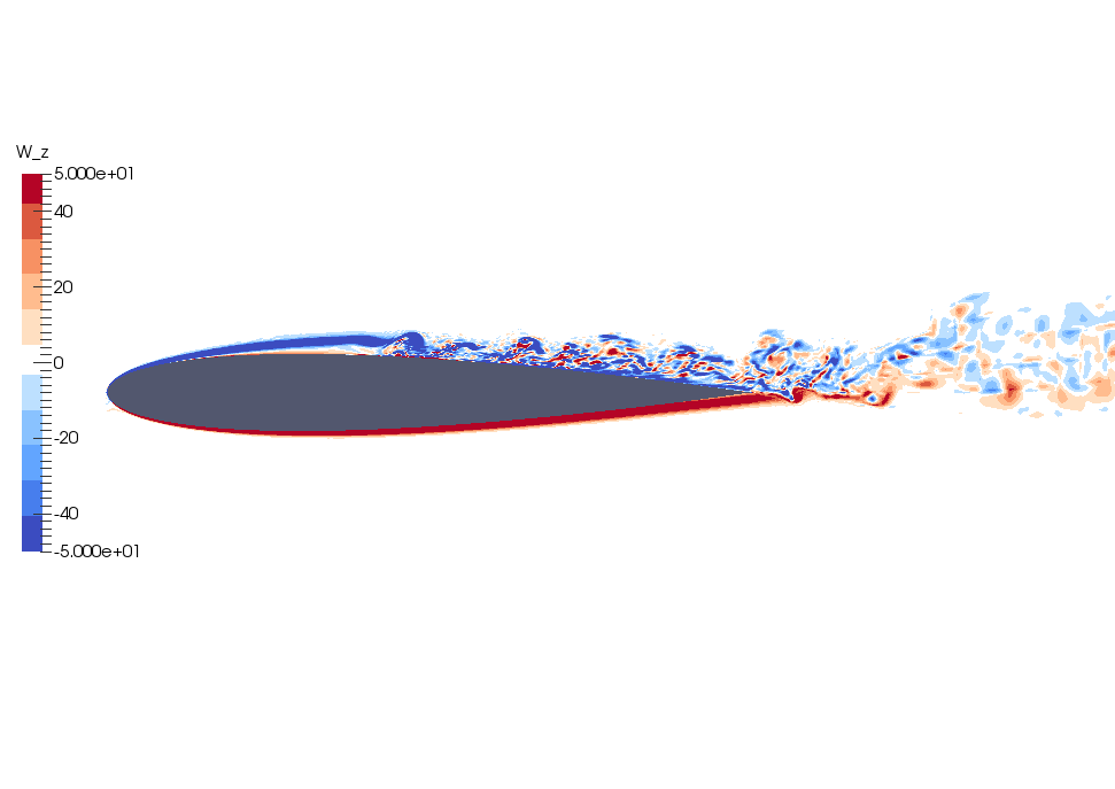}}}\hfill%
    \subcaptionbox{Polynomial order}[0.48\textwidth]{\fbox{\includegraphics[width=0.48\textwidth]{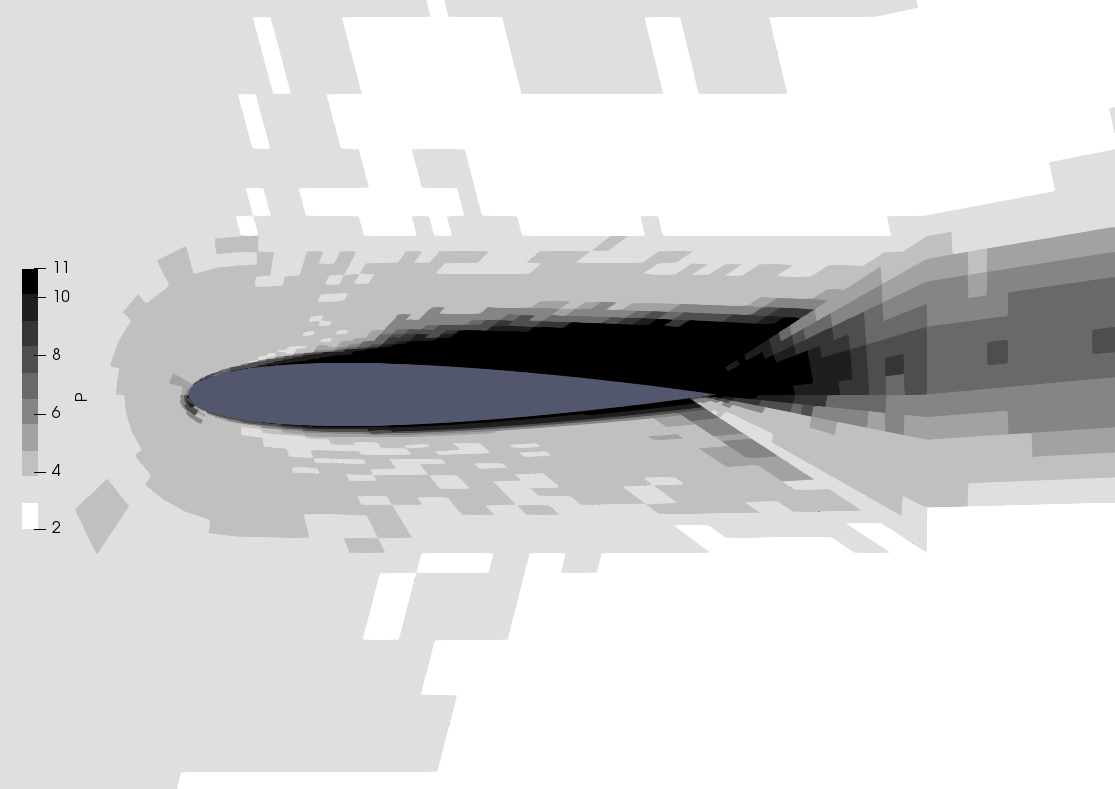}}}
    }%

    \caption{\label{fig:adaptiveP}Polynomial order and vorticity distributions
      obtained with simulation using adaptive polynomial order for the flow
      around a NACA 0012 profile with $Re=\numprint{50000}$ and
      $\alpha=6^{\circ}$. Taken from~\cite{nekpp-icosahom2016}.}
\end{figure*}

An example of an application of the adaptive polynomial order procedure is
presented in Figure~\ref{fig:adaptiveP}, showing the spanwise vorticity and
polynomial order distributions for a quasi-3D simulation of the incompressible
flow around a NACA 0012 profile at Reynolds number $Re=\numprint{50000}$ and
angle of attack $\alpha=6^{\circ}$. The session files to generate this data can
be found in Example~\ref{f:suppl:adaptive}. It is clear that the regions with
transition to turbulence and the boundary layers are resolved using the largest
polynomial order allowed, while regions far from the airfoil use a low
polynomial order. This way, the scheme succeeds in refining the discretisation
in the more critical regions where small scales are present, without incurring
in the large computational costs that would be required to uniformly increase
the polynomial order. More simply stated, it is possible to specify different
polynomial order in the quadrilateral elements (typically used in boundary layer
discretisation) and the triangle elements (typically used to fill the outer
volume). As a final point, we note that the use of variable polynomial order is
not limited to quasi-3D simulations; both CG and DG discretisations fully
support all element shape types in 2D and 3D, with parallel implementations
(including frequently used preconditioners) also supporting this discretisation
option.

\subsection{Global mapping}
\label{sec:mapping}
Even though the \shp element spatial discretization allows us to model complex
geometries, in some cases it can be advantageous to apply a coordinate
transformation for solving problems that lie in a coordinate system other than
the Cartesian frame of reference. This is typically the case when the
transformed domain possesses a symmetry; this allows us to solve the equations more
efficiently by compensating for the extra cost of applying the coordinate
transformation.  Examples of this occur when a transform can be used to map a non-homogeneous geometry to a homogeneous geometry in one or more directions.
This makes it possible to use the cheaper quasi-3D approach, where this direction is
discretized using a Fourier expansion, and also for problems with moving
boundaries, where we can map the true domain to a fixed computational domain, avoiding the need for recomputing the system matrices after every time-step.


The implementation of this method was achieved in two parts. First, a new
library called {\tt GlobalMapping} was created, implementing general tensor
calculus operations for several types of transformations. Even though it would
be sufficient to consider just a completely general transformation, specific
implementations for particular cases of simpler transformations are also
included in order to improve the computational efficiency, since in these
simpler cases many of the metric terms are trivial. In a second stage, the
incompressible Navier-Stokes solver was extended, using the functionality of the
{\tt GlobalMapping} library to implement the methods presented
in~\cite{SeMeSh2016}. Some examples of applications are given in~\cite{SeMeSh2017a,SeMeSh2017b}. Embedding these global mappings at the
library level allows similar techniques to be introduced in other solvers in the future.

\begin{figure}[htpb]
  \centering
  \includegraphics[width=0.48\textwidth]{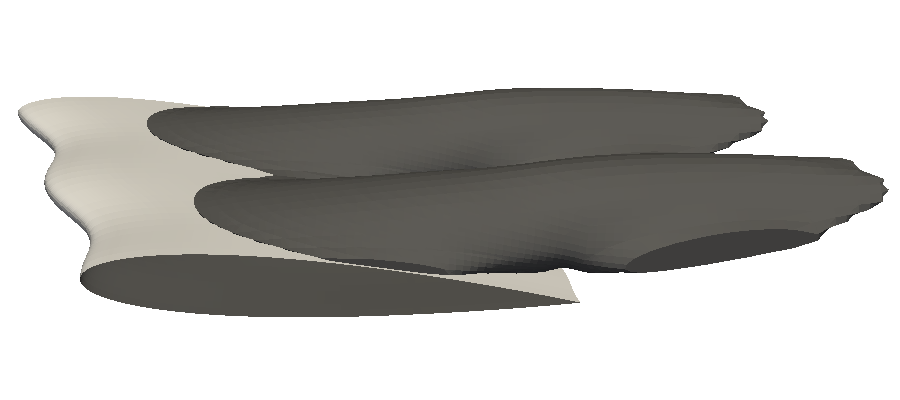}

  \caption{\label{fig:wavyWing}Time-averaged streamwise reversing regions for
    incompressible flow over a wing with spanwise waviness with
    $Re=\numprint{1000}$ and $\alpha=12^{\circ}$.}
\end{figure}

Figure~\ref{fig:wavyWing} presents an example of the application of
this technique, indicating the recirculation regions (i.e. regions
where the streamwise velocity is negative) for the flow over a wing
with spanwise waviness. In this case, the coordinate transformation
removes the waviness from the wing, allowing us to treat the
transformed geometry with the quasi-3D formulation. It is important to
note that this technique becomes unstable as the waviness amplitude
becomes too large. The fully explicit mapping is more sensitive to instability than
the semi-implicit mapping as discussed in~\cite{SeMeSh2016}. However, in cases
where it can be successfully applied, it leads to significant gains in
terms of computational cost when compared against a fully 3D
implementation. The session files used in this example can be found in
Example~\ref{f:suppl:wavy}.

\section{Applications}
\label{sec:applications}

In this section, we demonstrate some of the new features provided in our new
release, with a focus on application areas. 

\subsection{NekMesh}
\label{sec:nekmesh}
In the previous publication~\cite{nektarpp2015}, we briefly outlined the {\em
  MeshConvert} utility, which was designed to read various external file formats
and perform basic processing and format conversion. In the new release of \nek, {\em MeshConvert} has been
transformed into a new tool, called {\em NekMesh}, which provides a series of
tools for both the generation of meshes from an underlying CAD geometry, as well
as the manipulation of linear meshes to make them suitable for high-order
simulations.  While \emph{MeshConvert} was dedicated to the conversion of
external mesh file formats, the scope of \emph{NekMesh} has been significantly
broadened to become a true stand-alone high-order mesh generator.

The generation of high-order meshes in \emph{NekMesh} follows an \emph{a
  posteriori} approach proposed in~\cite{Peiro1}.  We first generate a linear
mesh using traditional low-order mesh generation techniques.  We then add
additional nodes along edges, faces and volumes to achieve a high-order
polynomial discretisation of our mesh. In the text below, we refer to these
additional nodes as `high-order' nodes, as they do not change the topology of
the underlying linear mesh, but instead deform it to fit a required geometry. In
this bottom-up procedure, these nodes are first added on edges, followed by
faces and finally the volume-interior.  At each step, nodes are generated on
boundaries to ensure a geometrically accurate representation of the model.

A key issue in this process, however, is ensuring that elements remain valid
after the insertion of high-order nodes, as this process is highly sensitive to
boundary curvature.  A common example of this is in boundary layer mesh
generation~\cite{MoHaPeSh13}, where elements are typically extremely thin in
order to resolve the high-shear of the flow near the wall. In this setting,
naively introducing curvature into the element will commonly push one face of
the element through another, leading a self-intersecting element and thus a mesh
that is invalid for computation.

\begin{figure}[htbp]
  \centering
  \includegraphics[width=0.45\textwidth]{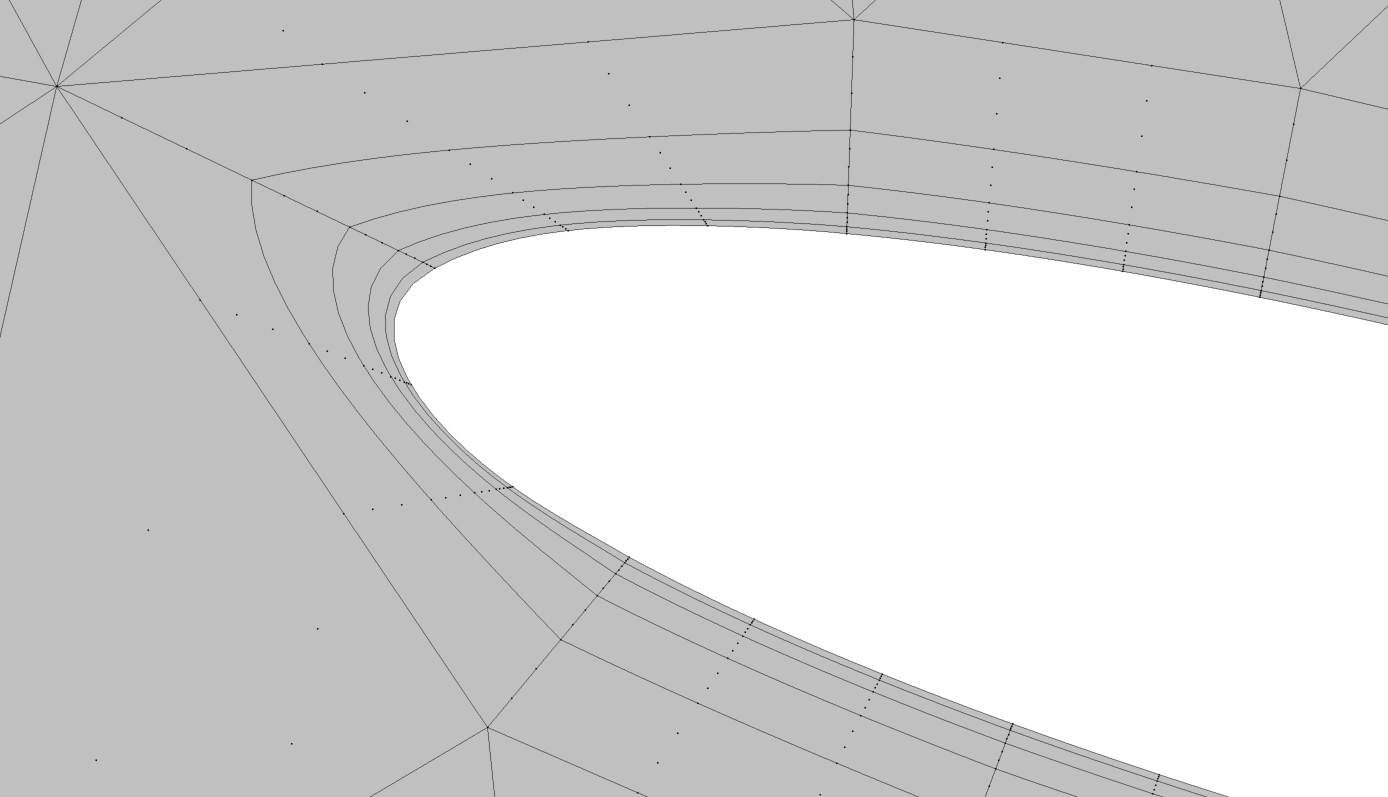}
  \caption{Example of split boundary layer mesh.}\label{fig:split}
\end{figure}

\begin{figure}[htbp]
  \centering
  \includegraphics[width=0.45\textwidth]{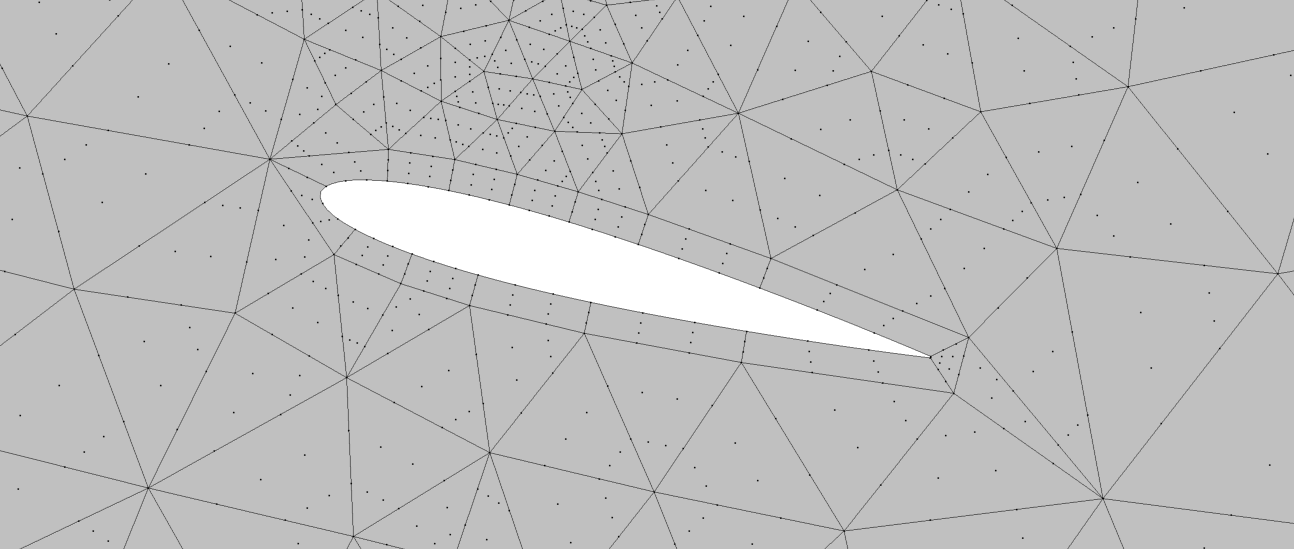}
  \caption{Example of mesh generated around a NACA 0012 aerofoil.}\label{fig:naca}
\end{figure}

\begin{figure}[htbp]
  \centering
  \includegraphics[width=0.45\textwidth]{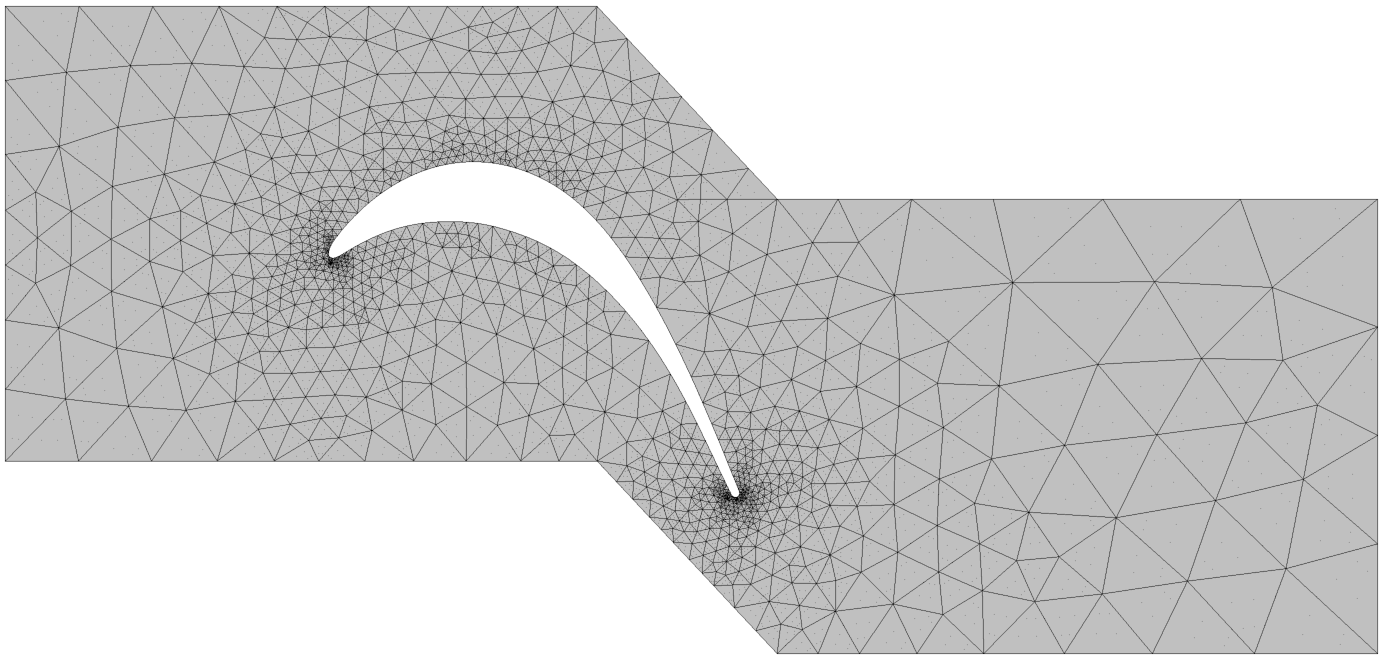}
  \caption{Example of mesh generated around a t106c turbine blade.}\label{fig:t106c}
\end{figure}

An important contribution of \emph{NekMesh} to the high-order mesh generation
community was presented in references~\cite{MoHaPeSh13,MoHaPeSh14}, where we
alleviate this risk through the creation of a coarse, single element boundary
layer mesh with edges orthogonal to the boundary.  The thickness of the layer of
elements gives enough room for a valid curving of the `macro'-elements.  After
creation of the high-order mesh, a splitting of these boundary elements can be
performed using the isoparametric mapping between the reference space and the
physical space. We then apply the original isoparametric mapping to construct
new elements within the `macro' element, thereby guaranteeing their
validity. This ensures conservation of the validity and quality of subdivided
elements while achieving very fine meshes.  An example is shown in
Figure~\ref{fig:split} where the coarse boundary layer mesh of
Figure~\ref{fig:naca} was split into five layers, using a geometric progression
of growth rate $r=2$ in the thickness of each layer. The session files used to
create the meshes for Figure~\ref{fig:naca} and~\ref{fig:t106c} can be found in
Example~\ref{f:suppl:meshgen}.

A complementary approach to avoid invalid or low quality high-order elements is
to optimise the location of high-order nodes in the mesh.  The approach proposed
in~\cite{turner2018curvilinear,Turner2017} of a variational framework for
high-order mesh optimisation was implemented in \emph{NekMesh}. In this
approach, we consider the mesh to be a solid body, and define a functional based
on the deformation of each high-order element. This functional can correspond to
physical solid mechanics governing equations such as linear or non-linear
elasticity, but also provides the possibility to accommodate arbitrary
functional forms such as the Winslow equations within the same framework.
Minimising this functional is then achieved through classical quasi-Newton
optimisation methods with the use of analytic gradient functions, alongside a
Jacobian regularisation technique to accomodate initially-invalid elements. As
demonstrated in~\cite{turner2018curvilinear}, the approach is scalable and
allows the possibility to implement advanced features, such as the ability to
slide nodes along a given constrained CAD curve or surface.

Along these lines, much of the development of \emph{NekMesh} has focused on the
access to a robust CAD system for CAD queries required for traditional meshing
operations. Assuming that the CAD is watertight, we note that only a handful of
CAD operations are required for mesh generation purposes. \emph{NekMesh}
therefore implements a lightweight wrapper around these CAD queries, allowing it
to be interfaced to a number of CAD systems. By default, we provide an interface
to the open-source OpenCASCADE library~\cite{OpenCascadeSAS2019}. OpenCASCADE is
able to read the STEP CAD file format, natively exported by most CAD design
tools, and load it into the system. At the same time, the use of a wrapper means
that users and developers of {\em NekMesh} are not exposed to the extensive
OpenCASCADE API. Although OpenCASCADE is freely available and very well suited
to simple geometries, it lacks many of the CAD healing tools required for more
complex geometries of the type typically found in industrial CFD environments,
which can frequently contain many imperfections and inconsistencies. However,
the use of a lightweight wrapper means that other commerical CAD packages can be
interfaced to {\em NekMesh} if available. To this end, we have implemented a
second CAD interface to the commerical CFI CAD engine, which provides a highly
robust interface and is described further in
references~\cite{Marcon2018,Marcon2019,Turner2017}.

While users are recommended to create their CAD models in a dedicated CAD
software, export them in STEP format and load them in \emph{NekMesh}, they also
have the possibility to create their own simple two-dimensional models using one
of two tools made available to them.  The first tool is an automatic NACA
aerofoil generator.  With just three inputs -- a NACA code, an angle of attack
and dimensions of the bounding box -- a geometry is generated and passed to the
meshing software.  An example is shown in Figure~\ref{fig:naca} of a mesh
generated around a NACA 0012 aerofoil at an angle of attack of
$\alpha = 15^{\circ}$.

The other tool is based on the GEO geometry file format of the
Gmsh~\cite{Geuzaine2009} open source mesh generator.  The GEO format is an
intepreter for the procedural generation of geometries.  \emph{NekMesh} has been
made capable to understand basic GEO commands, which gives the possibility to
generate simple two-dimensional geometries. An example is shown in
Figure~\ref{fig:t106c} of a mesh generated around a T106C turbine blade: the
geometry was created using a GEO script of lines and splines.

\subsection{Acoustic solver}
\label{sec:acoustic}
%
Time-domain computational aeroacoustics simulations are commonly used to model
noise emission over wide frequency ranges or to augment flow simulations in
hybrid setups.  Compared with fully compressible flow simulations, they require
less computational effort due to the reduced complexity of the governing
equations and larger length scales~\cite{Colonius2004}.  However, due to the
small diffusive terms, as well as the long integration times and distances
required for these simulations, highly accurate numerical schemes are crucial
for stability~\cite{Tam2006}. This numerical accuracy can be provided by \shp
element methods, even on unstructured meshes in complex geometries, and hence
\nek provides a good candidate framework on which to build such an application
code.

%
The latest release of \nek includes a new {\tt AcousticSolver}, which implements
several variants of aeroacoustic models. These are formulated in a hyperbolic
setting and implemented in a similar fashion to the compressible Euler and
Navier-Stokes equations, encapsulated in \nek inside the {\tt
  CompressibleFlowSolver}.  Following this implementation guideline, the {\tt
  AcousticSolver} uses a discontinuous Galerkin spatial discretisation with
modal or nodal expansions to model time-domain acoustic wave propagation in one,
two or three dimensions.  It implements the operators of the linearized Euler
Equations (LEE) and the Acoustic Perturbation Equations 1 and 4
(APE-1/4)~\cite{Ewert2003}, both of which describe the evolution of
perturbations around a base flow state.
For the APE-1/4 operator, the system is defined by the hyperbolic equations
\begin{subequations}
    \begin{align}
    \frac{\partial p^\mathrm{a}}{\partial t} 
    + \overline{c^2} \nabla \cdot \left( \overline{\rho} \bm{u}^\mathrm{a} 
    + \overline{\bm{u}} \frac{p^\mathrm{a}}{\overline{c^2}} \right) 
    &= \dot{\omega}_\mathrm{c},
    \\
    \frac{\partial \bm{u}^\mathrm{a}}{\partial t} 
    + \nabla \left(\overline{\bm{u}} \cdot \bm{u}^\mathrm{a} \right) 
    + \nabla \left(\frac{p^\mathrm{a}}{\overline{\rho}}\right) 
    &= \dot{\bm{\omega}}_\mathrm{m},
    \end{align}
    \label{eq:APE}
\end{subequations}
where $\bm{u}$ denotes the flow velocity, $\rho$ its density, $p$ its pressure
and $c$ corresponds to the speed of sound. The quantities $\bm{u}^a$ and $p^a$
refer to the irrotational acoustic perturbation of the flow velocity and its
pressure, with overline quantities such as $\overline{\bm{u}}$ denoting the
time-averaged mean. The right-hand-side acoustic source terms terms
$\dot{\omega}_\mathrm{c}$ and $\dot{\boldsymbol{\omega}}_\mathrm{m}$ are
specified in the session file.  This allows for the implementation of any
acoustic source term formulation so that, for example, the full APE-1 or APE-4
can be obtained. In addition to using analytical expressions, the source terms
and base flow quantities can be read from disk or transferred from coupled
applications, enabling co-simulation with a driving flow solver.
Both, LEE and APE support non-quiescent base flows with inhomogeneous speed of
sounds.  Accordingly, the Lax-Friedrichs and upwind Riemann solvers used in the
{\tt AcousticSolver} employ a formulation which is valid even for strong base
flow gradients.
The numerical stability can be further improved by optional sponge layers and
suitable boundary conditions, such as rigid wall, farfield or white noise.

\begin{figure}[htb]
  \centering
  \begin{subfigure}[c]{\linewidth}
    \centering
    \footnotesize    
    \def\svgwidth{0.7\linewidth}
    \begingroup%
    \makeatletter%
    \providecommand\color[2][]{%
      \errmessage{(Inkscape) Color is used for the text in Inkscape, but the package 'color.sty' is not loaded}%
      \renewcommand\color[2][]{}%
    }%
    \providecommand\transparent[1]{%
      \errmessage{(Inkscape) Transparency is used (non-zero) for the text in Inkscape, but the package 'transparent.sty' is not loaded}%
      \renewcommand\transparent[1]{}%
    }%
    \providecommand\rotatebox[2]{#2}%
    \newcommand*\fsize{\dimexpr\f@size pt\relax}%
    \newcommand*\lineheight[1]{\fontsize{\fsize}{#1\fsize}\selectfont}%
    \ifx\svgwidth\undefined%
    \setlength{\unitlength}{276.76206333bp}%
    \ifx\svgscale\undefined%
    \relax%
    \else%
    \setlength{\unitlength}{\unitlength * \real{\svgscale}}%
    \fi%
    \else%
    \setlength{\unitlength}{\svgwidth}%
    \fi%
    \global\let\svgwidth\undefined%
    \global\let\svgscale\undefined%
    \makeatother%
    \begin{picture}(1,0.84634509)%
      \lineheight{1}%
      \setlength\tabcolsep{0pt}%
      \put(0,0){\includegraphics[width=\unitlength,page=1]{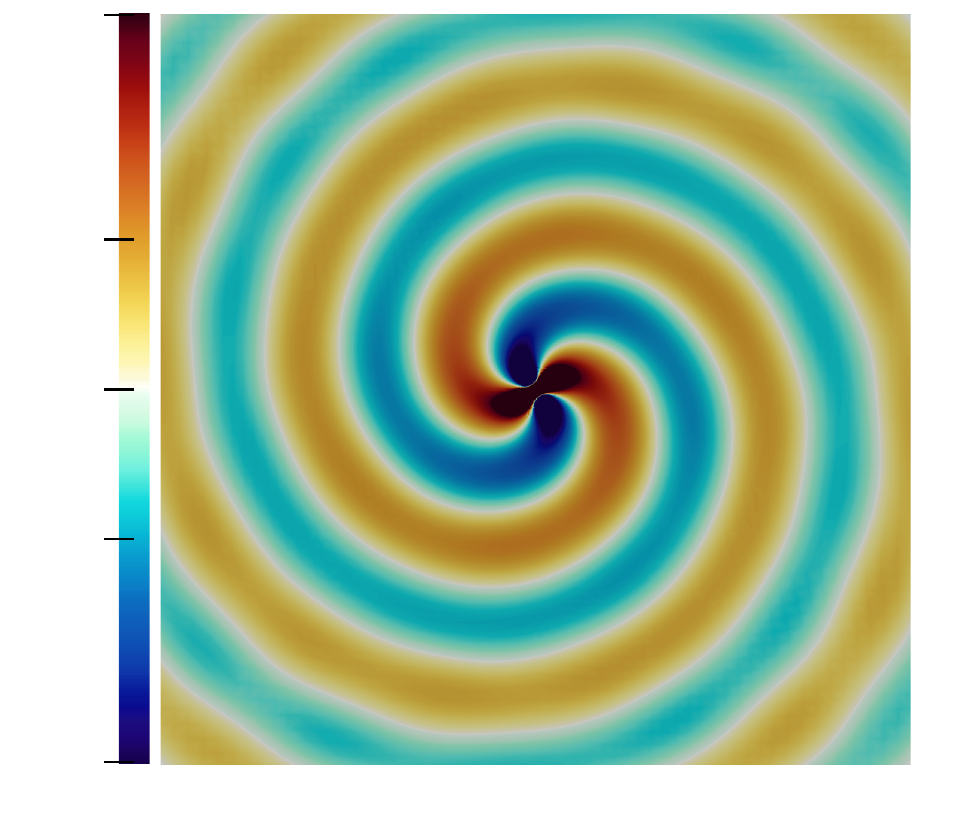}}%
      \put(0.10254606,0.03763946){\color[rgb]{0,0,0}\makebox(0,0)[rt]{\lineheight{0}\smash{\begin{tabular}[t]{r}-5E-4\end{tabular}}}}%
      \put(0.10254606,0.81737058){\color[rgb]{0,0,0}\makebox(0,0)[rt]{\lineheight{0}\smash{\begin{tabular}[t]{r}5E-4\end{tabular}}}}%
      \put(0.10254606,0.27187721){\color[rgb]{0,0,0}\makebox(0,0)[rt]{\lineheight{0}\smash{\begin{tabular}[t]{r}-2E-4\end{tabular}}}}%
      \put(0.10254606,0.58131899){\color[rgb]{0,0,0}\makebox(0,0)[rt]{\lineheight{0}\smash{\begin{tabular}[t]{r}2E-4\end{tabular}}}}%
      \put(0.1029536,0.42547612){\color[rgb]{0,0,0}\makebox(0,0)[rt]{\lineheight{0}\smash{\begin{tabular}[t]{r}0\end{tabular}}}}%
      \put(0.03615681,0.44001188){\color[rgb]{0,0,0}\rotatebox{90}{\makebox(0,0)[t]{\lineheight{0}\smash{\begin{tabular}[t]{c}$p^\mathrm{a}/p$ [-]\end{tabular}}}}}%
      \put(0,0){\includegraphics[width=\unitlength,page=2]{vortex.pdf}}%
      \put(0.55547514,0.00937349){\color[rgb]{0,0,0}\makebox(0,0)[t]{\lineheight{0}\smash{\begin{tabular}[t]{c}$x_1/r_0$ [-]\end{tabular}}}}%
      \put(0.99062644,0.44159125){\color[rgb]{0,0,0}\rotatebox{90}{\makebox(0,0)[t]{\lineheight{0}\smash{\begin{tabular}[t]{c}$x_2/r_0$ [-]\end{tabular}}}}}%
      \put(0,0){\includegraphics[width=\unitlength,page=3]{vortex.pdf}}%
    \end{picture}%
    \endgroup%
    \caption{Normalized acoustic pressure distribution at $t = 1\text{s}$ with
      the mesh shown in light gray and the sampling line in a black dashed
      line.}
    \label{fig:vortex}
  \end{subfigure}
  \\
  \begin{subfigure}[c]{\linewidth}
    \centering
    \footnotesize
    \def\svgwidth{0.7\linewidth}
    \begingroup%
    \makeatletter%
    \providecommand\color[2][]{%
      \errmessage{(Inkscape) Color is used for the text in Inkscape, but the package 'color.sty' is not loaded}%
      \renewcommand\color[2][]{}%
    }%
    \providecommand\transparent[1]{%
      \errmessage{(Inkscape) Transparency is used (non-zero) for the text in Inkscape, but the package 'transparent.sty' is not loaded}%
      \renewcommand\transparent[1]{}%
    }%
    \providecommand\rotatebox[2]{#2}%
    \newcommand*\fsize{\dimexpr\f@size pt\relax}%
    \newcommand*\lineheight[1]{\fontsize{\fsize}{#1\fsize}\selectfont}%
    \ifx\svgwidth\undefined%
    \setlength{\unitlength}{218.2461853bp}%
    \ifx\svgscale\undefined%
    \relax%
    \else%
    \setlength{\unitlength}{\unitlength * \real{\svgscale}}%
    \fi%
    \else%
    \setlength{\unitlength}{\svgwidth}%
    \fi%
    \global\let\svgwidth\undefined%
    \global\let\svgscale\undefined%
    \makeatother%
    \begin{picture}(1,1.00830484)%
      \lineheight{1}%
      \setlength\tabcolsep{0pt}%
      \put(0.18944301,0.10352303){\color[rgb]{0,0,0}\makebox(0,0)[t]{\lineheight{0}\smash{\begin{tabular}[t]{c}0\end{tabular}}}}%
      \put(0.44352814,0.10352303){\color[rgb]{0,0,0}\makebox(0,0)[t]{\lineheight{0}\smash{\begin{tabular}[t]{c}50\end{tabular}}}}%
      \put(0,0){\includegraphics[width=\unitlength,page=1]{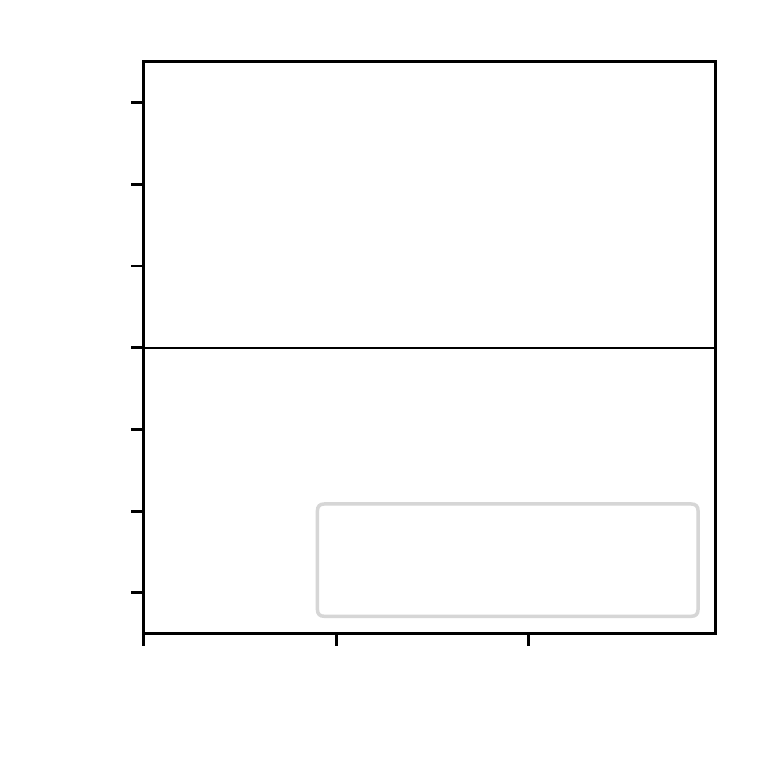}}%
      \put(0.06941049,0.54742651){\color[rgb]{0,0,0}\rotatebox{90}{\makebox(0,0)[t]{\lineheight{0}\smash{\begin{tabular}[t]{c}$p^\mathrm{a}/p$ [-]\end{tabular}}}}}%
      \put(0.69756222,0.10352303){\color[rgb]{0,0,0}\makebox(0,0)[t]{\lineheight{0}\smash{\begin{tabular}[t]{c}100\end{tabular}}}}%
      \put(0.1545616,0.53107045){\color[rgb]{0,0,0}\makebox(0,0)[rt]{\lineheight{0}\smash{\begin{tabular}[t]{r}0\end{tabular}}}}%
      \put(0.15586594,0.42323573){\color[rgb]{0,0,0}\makebox(0,0)[rt]{\lineheight{0}\smash{\begin{tabular}[t]{r}-1\end{tabular}}}}%
      \put(0.1562597,0.31509339){\color[rgb]{0,0,0}\makebox(0,0)[rt]{\lineheight{0}\smash{\begin{tabular}[t]{r}-2\end{tabular}}}}%
      \put(0.15525068,0.20764013){\color[rgb]{0,0,0}\makebox(0,0)[rt]{\lineheight{0}\smash{\begin{tabular}[t]{r}-3\end{tabular}}}}%
      \put(0.15586594,0.63885594){\color[rgb]{0,0,0}\makebox(0,0)[rt]{\lineheight{0}\smash{\begin{tabular}[t]{r}1\end{tabular}}}}%
      \put(0.1562597,0.74633379){\color[rgb]{0,0,0}\makebox(0,0)[rt]{\lineheight{0}\smash{\begin{tabular}[t]{r}2\end{tabular}}}}%
      \put(0.15525068,0.85450073){\color[rgb]{0,0,0}\makebox(0,0)[rt]{\lineheight{0}\smash{\begin{tabular}[t]{r}3\end{tabular}}}}%
      \put(0.19051988,0.93878397){\color[rgb]{0,0,0}\makebox(0,0)[lt]{\lineheight{0}\smash{\begin{tabular}[t]{l}1E-4\end{tabular}}}}%
      \put(0.56573388,0.29119648){\color[rgb]{0,0,0}\makebox(0,0)[lt]{\lineheight{1.25}\smash{\begin{tabular}[t]{l}Analytical\end{tabular}}}}%
      \put(1.53735632,-0.00562665){\color[rgb]{0,0,0}\makebox(0,0)[lt]{\begin{minipage}{0.192777\unitlength}\raggedright \end{minipage}}}%
      \put(0.56554867,0.22520724){\color[rgb]{0,0,0}\makebox(0,0)[lt]{\lineheight{1.25}\smash{\begin{tabular}[t]{l}AcousticSolver\end{tabular}}}}%
      \put(0,0){\includegraphics[width=\unitlength,page=2]{vortex_line.pdf}}%
      \put(0.56546939,0.03726182){\color[rgb]{0,0,0}\makebox(0,0)[t]{\lineheight{1.25}\smash{\begin{tabular}[t]{c}$r / r_0$ [-]\end{tabular}}}}%
    \end{picture}%
    \endgroup%
    \caption{Normalized acoustic pressure along sample line, obtained with the
      AcousticSolver and analytical solution \cite{Muller1967}.}
    \label{fig:vortex:line}
  \end{subfigure}
  \caption{Normalized acoustic pressure for $\Gamma/(c r_0) = 1.0$ and
    $\mathit{Ma}_r = 0.0795$ at $t =1\mathrm{s}$.}
\end{figure}

%
A recurring test case for APE implementations is the ``spinning vortex
pair''~\cite{Muller1967}.  It is defined using two opposing vortices, that are
each located at $r_0$ from the center $x_1 = x_2 = 0$ of a square domain with
edge length $-100 \, r_0 \leq x_{1,2} \leq 100 \, r_0$.
The vortices have a circulation of $\Gamma$ and rotate around the center at the
angular frequency $\omega = \Gamma / 4 \pi r_0^2$ and circumferential Mach
number $\mathit{Ma}_r = \Gamma / 4 \pi r_0 c$.
The resulting acoustic pressure distribution is shown in Figure~\ref{fig:vortex}
and was obtained on an unstructured mesh of \numprint{465} quadrilateral
elements with a fifth order modal expansion ($P=5$). The session files used to
generate this example can be found in Example~\ref{f:suppl:acoustic}.
Along the black dashed line, the acoustic pressure shown in
Figure~\ref{fig:vortex:line} exhibits minor deviations from the analytical
solution defined in~\cite{Muller1967}, but is in excellent agreement with the
results of the original simulation in \cite{Ewert2003}.  The latter was based on
a structured mesh with \numprint{19881} nodes and employed a sponge layer
boundary condition and spatial filtering to improve the stability.  Due to the
flexibility and numerical accuracy of the \shp method, a discretization with
only \numprint{16740} degrees of freedom was sufficient for this simulation, and
no stabilization measures (e.g. SVV or filtering) were necessary to reproduce
this result.

\subsection{Fluid-Structure Interaction (FSI) and Vortex-Induced Vibration (VIV)}
\label{sec:viv}
Fluid-structure interaction (FSI) modelling poses a great challenge for the accurate
prediction of vortex-induced vibration (VIV) of long flexible bodies, as the
full resolution of turbulent flow along their whole span requires considerable computational resources. This is particularly true in the
case of large aspect-ratio bodies. Although 2D strip-theory-based modelling of
such problems is much more computationally efficient, this approach is unable to
resolve the effects of turbulent fluctuations on dynamic coupling of FSI
systems~\cite{ChBe05,WiGr01}. A novel strip model, which we refer to as `thick'
strip modelling, has been developed using the the \nek framework
in~\cite{bao2016generalized}, whose implementation is supported within the
incompressible Navier-Stokes solver. In this approach, a three-dimensional DNS
model with a local spanwise scale is constructed for each individual
strip. Coupling between strips is modelled implicitly through the structural
dynamics of the flexible body.

In the `thick' strip model, the flow dynamics are governed by a series of
incompressible Navier-Stokes equations. The governing equations over a general
local domain $\mathit\Omega _n$ associated with the $n$-th strip are written as
\begin{equation}\label{eq:ns-0}
  {\partial{\mathbf u_n}\over{\partial t}}
  +({\mathbf u_n} \cdot \nabla) {\mathbf u_n} 
  = - \nabla p_n +  {1\over Re}\nabla^2
  {\mathbf u_n} \quad \mbox{on}  \quad \mathit \Omega _n
\end{equation}
%
\begin{equation}\label{eq:ns-1}
	\nabla\cdot{\mathbf u_n} = 0 
	\quad \mbox{on}  \quad \mathit \Omega _n,
\end{equation}
where the vector $\mathbf{u}_n=(u_n,v_n,w_n)$ denotes the fluid velocity inside
the $n$-th strip, with $p_n$ being the corresponding dynamic pressure and $Re$
the Reynolds number, which we assume to be constant across all strips. The
governing equations are supplemented by boundary conditions of either Dirichlet
or Neumann type. In particular, no-slip boundary conditions are applied to the
wall of the body, and the velocity of the moving wall is imposed and determined
from the transient solution of structural dynamics equations of motion. A
linearized tensioned beam model is used to describe the structural dynamic
behavior of the flexible body, which is expressed by the system
%
\begin{equation}\label{eq:ns-2}
	\rho_c{\partial^{2}{\bm \eta}\over{\partial t}^{2}}
	+{c}{\partial{\bm \eta}\over{\partial t}}
	-{T}{\partial^{2}{\bm \eta}\over{\partial z}^{2}}
	+{EI}{\partial^{4}{\bm \eta}\over
	{\partial z}^{4}}={\bf f}.
\end{equation}
In the above, $\rho_{c}$ is the structural mass per unit length, $c$ is the
structural damping per unit length, $T$ is the tension and $EI$ is the flexural
rigidity. $\mathbf{f}$ denotes the vector of hydrodynamic force per unit length
exerted on the body's wall and $\bm{\eta}$ is the structural displacement
vector.

Homogeneity is imposed in the spanwise direction to the local flow within
individual strips, under the assumption that the width of the strips is much
shorter with respect to the oscillation wavelength of excited higher-order modes
of the flexible body. This therefore enables the use of the
computationally-efficient quasi-3D approach discussed in previous sections
within each strip domain, in which two-dimensional spectral elements with
piecewise polynomial expansions are used in the $(x,y)$ plane and Fourier
expansions are used in the homogeneous $z$ direction. This also requires the
assumption of a spanwise oscillation of the flexible body with respect to its
full-length scale. As a consequence, the motion variables and fluid forces are
expressed as a complex Fourier series, and the tensioned beam model is decoupled
into a set of ordinary differential equations, which can be solved simply by a
second-order Newmark-$\beta$ method~\cite{NeKa97}. A partitioned approach is
adopted to solve the coupled FSI system, in which coordinate mapping technique
discussed in Section~\ref{sec:mapping} is implemented for the treatment of the
moving wall~\cite{SeMeSh2016}.

\begin{figure}[!htb]
  \centering
  \includegraphics[width=9.0cm]{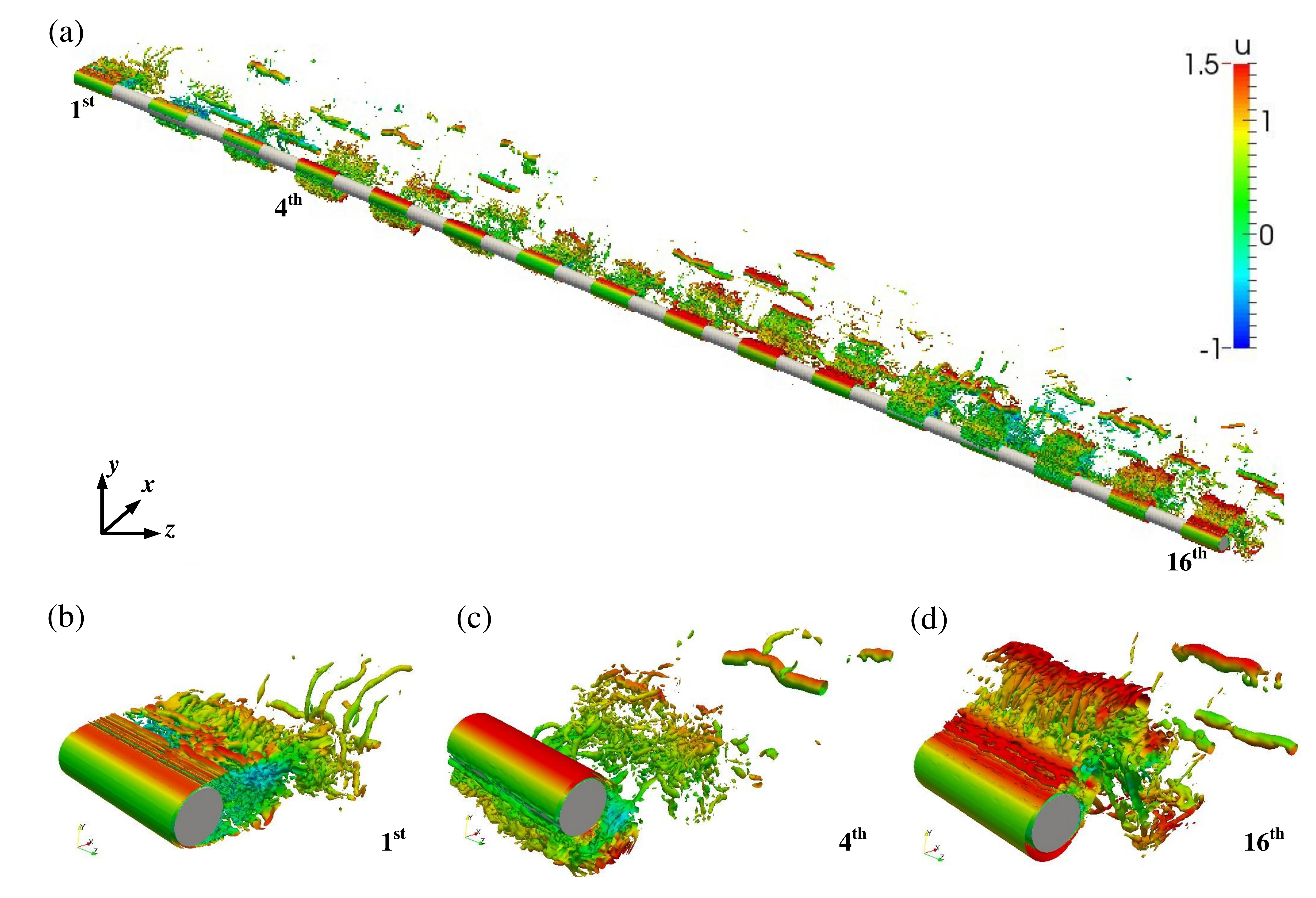}
  \caption{Instantaneous vortex shedding visualized by the vortex-identification
    of Q-criterion (iso-surfaces of the Q-value= $[-5, 5]$) in body-fitted
    coordinates: (a) full domain view and zoom-in view of (b) first strip; (c)
    fourth strip and (d) 16th strip.}
  \label{fig:VIVofFlexCyl} 
\end{figure}

To illustrate the application of this modelling approach, VIV of a long flexible
cylinder with an aspect ratio of $32\pi$ which is pinned at both ends is
simulated at $Re=\numprint{3900}$, with 16 thick strips allocated evenly along
the axial line of the cylinder. The instantaneous spanwise wake structure is
visualized by the vortex-identification of Q-criterion in
Figure~\ref{fig:VIVofFlexCyl}. As the figure demonstrates, the distribution of
vortex shedding illustrates that a second harmonic mode is excited along the
spanwise length and the turbulent structure is captured well in the local domain
of the strips. This emphasises the convincing advantage of providing
highly-resolved description of hydrodynamics involved in the FSI process. The
session files used to run this simulation can be found in
Example~\ref{f:suppl:viv}.

\subsection{Aeronautical applications}
\label{sec:aero}

CFD is now an indispensable tool for the design of aircraft engines, and it has
become commonplace in the design guidance of new technologies and
products~\cite{Laskowski2016}.  In order for CFD to be effectively adopted in
industry, validation and verification is required over a broad design
space. This is challenging for a number of reasons, including the range of
operating conditions (i.e. Reynolds numbers, Mach numbers, temperatures
and pressures), the complexity of industrial geometries (including uncertainty
due to manufacturing variations) and their relative motion
(i.e. rotor-stator interactions).  Even though RANS continues to be the backbone of CFD-based design,
the recent development of high-order unstructured solvers and high-order
unstructured meshing algorithms, combined with the lowering cost of HPC
infrastructures, has the potential to allow for the introduction of
high-fidelity transient simulations using large-eddy or direct numerical
simulations (LES/DNS) in the design loop, taking the role of a virtual wind
tunnel.

As part of our effort to bridge the gap between academia and industry, we have
been developing the expertise to analyse turbomachinery cases of industrial
interest using \nek. A key problem to overcome in these cases is the sensitivity
of these simulations to variations in spatial resolution, which requires the use
of stabilisation techniques in order to improve robustness. \nek makes use of
the spectral vanishing viscosity (SVV) method, originally introduced for the
Fourier spectral method by Tadmor~\cite{tadmor1989convergence}. SVV is a
model-free stabilization technique that acts at the subgrid-scale level and
allows for exponential convergence properties to be conserved in sufficiently
resolved simulations. Recent developments in this area have focused on a new SVV
kernel by Moura et al.~\cite{mengaldo2018spatial_a}, which replicates the desirable
dispersion and diffusion properties of DG schemes and does not require the
manual tuning of parameters found in the classical SVV formulation. More
specifically, the dissipation curves of the CG scheme of order $P$ were compared
to those of DG order $P-2$, and the DG kernel was determined from minimization
of the point-wise $L_2$ norm between these curves. SVV stabilization is combined
with spectral/\emph{hp} dealiasing~\cite{mengaldo2015dealiasing} to eliminate
errors arising from the integration of non-linear terms.

\begin{figure}[!htb]
  \centering
  \includegraphics[width=9.0cm]{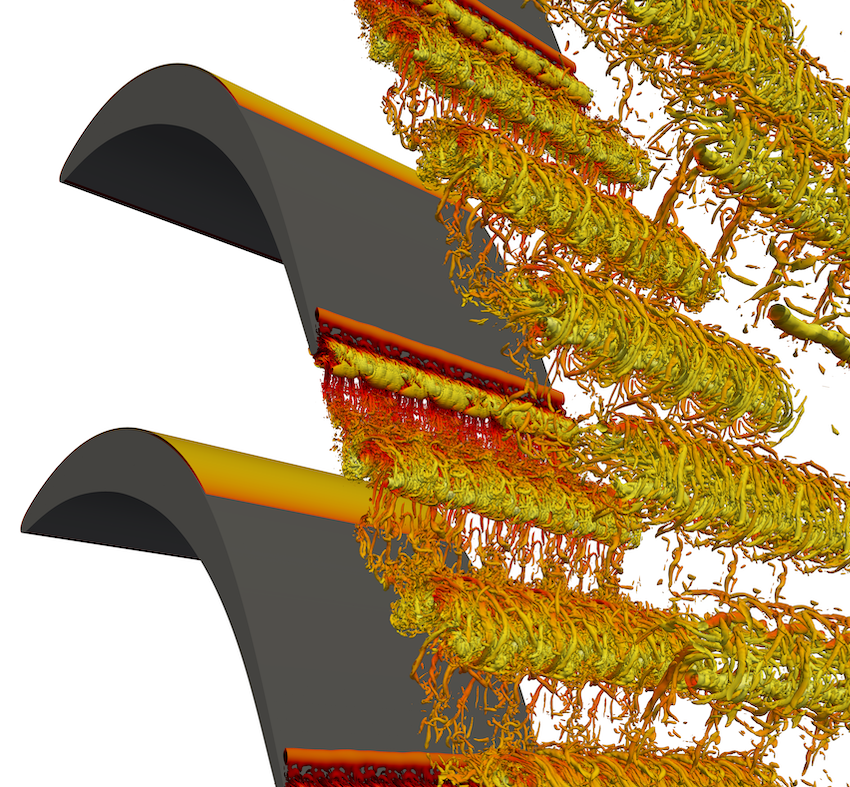}
  \caption{Instantaneous isosurfaces of $Q$-criterion ($Q=500$) contoured by
    velocity magnitude, showing the vortical structures evolving on the suction
    surface and in the wake of a T106A cascade. The computational domain is
    replicated in the spanwise and pitchwise directions for visual clarity.}
  \label{fig:LPT_Qiso}
\end{figure}

A T106A low pressure turbine vane was investigated at moderate regime
($Re=\numprint{88450}$), and the convergence properties of the main flow
statistics were extensively explored with the aim of developing a set of best
practices for the use of spectral/\emph{hp} element methods as a high-fidelity
digital twin~\cite{Cassinelli2018}.  The velocity correction scheme
of~\cite{KaIsOr91} implemented in the {\tt IncNavierStokesSolver} is adopted,
using the quasi-3D approach discussed in the previous sections and Taylor-Hood
type elements in 2D (where spaces of order $P$ polynomials on each element are
used for the velocity components, and $P-1$ for pressure).  Uniform inflow
velocity is combined with pitchwise periodicity and high-order outflow boundary
conditions~\cite{DoKaCh14}. Numerical stability is ensured by employing SVV with
the DG kernel in the $x$-$y$ planes, and the traditional exponential kernel for
the spanwise Fourier direction. A representation of the vortical structures is
shown in Figure~\ref{fig:LPT_Qiso}: transition to turbulence takes place only in
the final portion of the suction surface, where the separated shear layer rolls
up due to Kevin-Helmoltz instability. The separation bubble remains open and
merges into the trailing edge wake, giving rise to large-scale vortical
structures. This work was conducted with clean inflow conditions to isolate the
effect of the numerical setup on the various flow statistics.  However,
turbomachinery flows are highly turbulent: subsequent work focused on the
treatment of flow disturbances to reproduce more accurately a realistic
environment~\cite{Cassinelli2019}.  With this aim, a localised synthetic
momentum forcing was introduced in the leading edge region to cause flow
symmetry breakdown on the suction surface, and promote anticipated transition to
turbulence.  This approach yields an improvement in the agreement with
experimental data, with no increase in the computational cost.

\begin{figure}
  \centering
  \includegraphics[width=1.0\columnwidth]{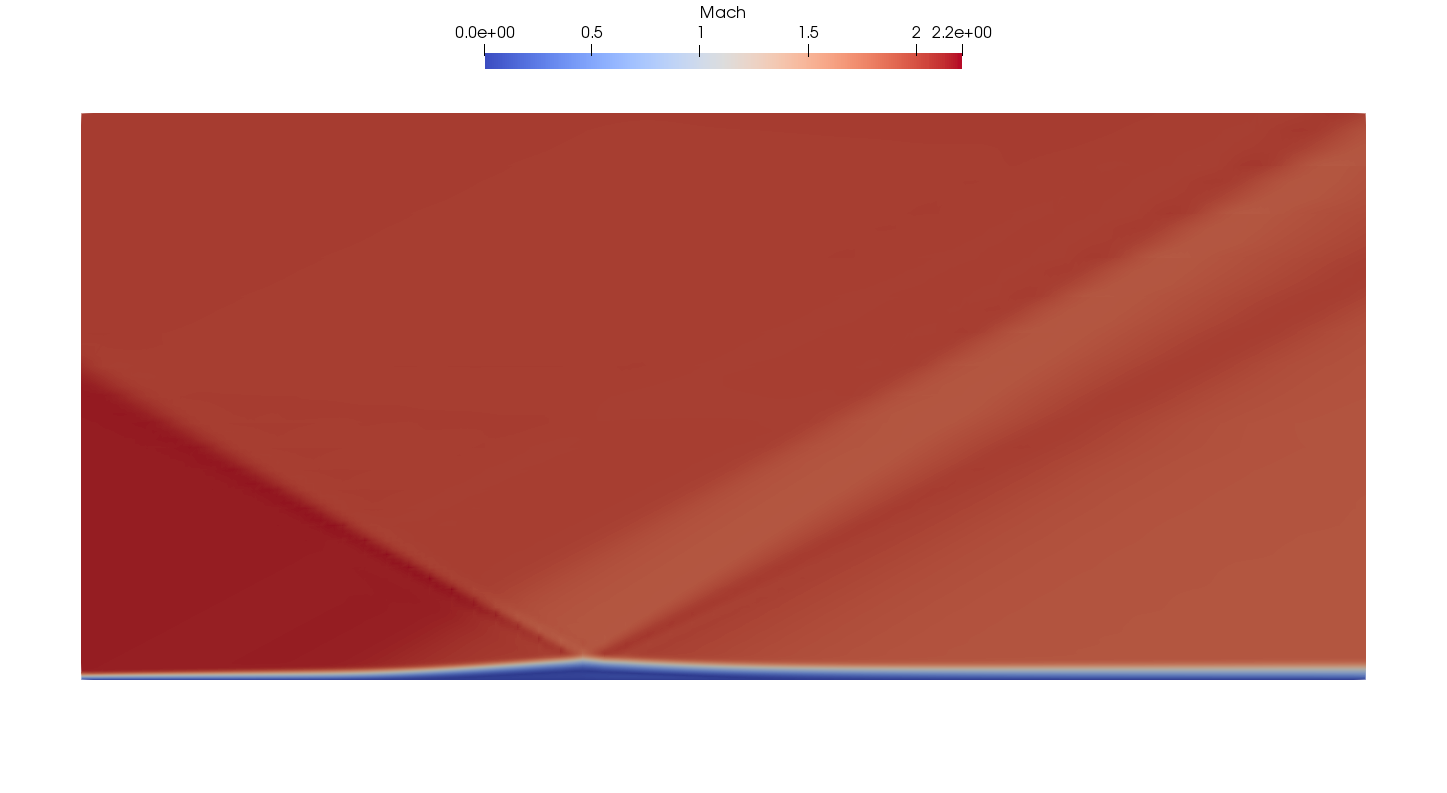}
  \caption{Mach number field of SWBLI test case ($60 \times 40$ quadrilateral elements, $p=3$); configuration based on \cite{degrez1987interaction}.}
  \label{fig:app-aero-swbli-mach}
\end{figure}

\begin{figure}
  \centering
  \includegraphics[width=1.0\columnwidth]{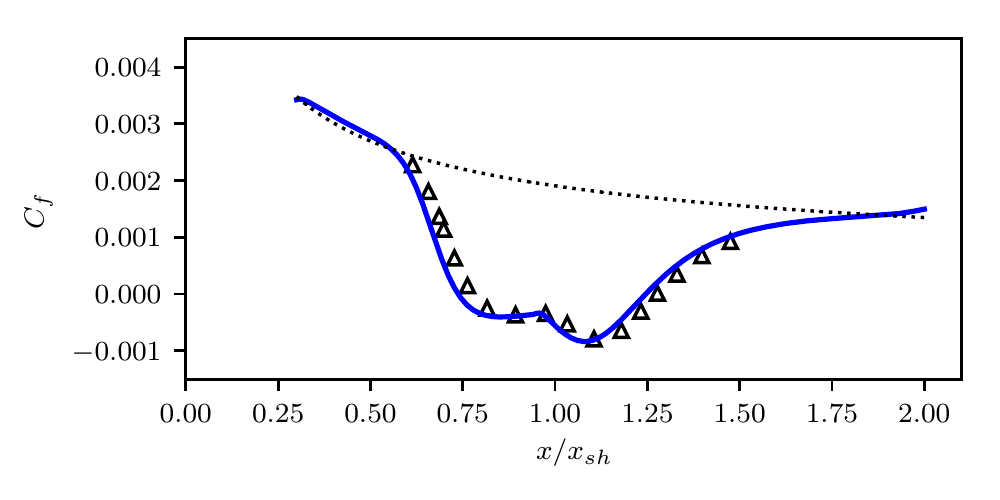}
  \caption{Skin friction coefficient for the SWBLI test case: blue line \nek ($60 \times 40$ quadrilateral elements, $p=3$); triangles are from \cite{boin20063d}; dotted line is empirical solution by \cite{eckert1955engineering}.}
  \label{fig:app-aero-swbli-cf}
\end{figure}

With the intent of being able to tackle cases in which compressibility effects
are not negligible, there has been an effort in validating the {\tt
  CompressibleFlowSolver} for shock-wave boundary layer interaction (SWBLI)
configurations. This solver, described in our previous
publication~\cite{nektarpp2015}, formulates the compressible Navier-Stokes
equations in their conservative form, discretised using a DG scheme and explicit
timestepping methods. In order to regularize the solution in the presence of
discontinuities, the right hand side of the Navier-Stokes equations is augmented
with a Laplacian viscosity term of the form
$\nabla \cdot \left( \varepsilon \nabla \mathbf{q} \right)$, where $\mathbf{q}$
is the vector of conserved variables, and $\varepsilon$ is a spatially-dependent
diffusion term that is defined on each element as
\begin{equation*}
  \varepsilon = \varepsilon_0 \frac{h}{p} \lambda_{\text{max}} S.
\end{equation*}
Here, $\varepsilon_0$ is a $O(1)$ constant, $\lambda_{\text{max}}$ is the
maximum local characteristic speed, $h$ is a reference length of the element,
$p$ its polynomial order, and $S$ a discontinuity sensor value using the
formulation of~\cite{PePe2006}. To benchmark this approach in the context of
SWBLI problems, we consider a laminar problem studied experimentally and
numerically in~\cite{degrez1987interaction}.  Several authors have studied this
SWBLI with slightly different free stream conditions; here we follow the
physical parameters used by~\cite{boin20063d}, where we select a free-stream
Mach number $Ma=2.15$, shock angle $\beta=30.8^\circ$, a stagnation pressure
$p_0=1.07\times10^4$ Pa, a stagnation temperature of $T_0=293 K$, a Reynolds
number $Re=10^5$ (referred to the inviscid shock impingement location
$x_{\text{sh}}$ measured from the plate leading edge), and a Prandtl number
$Pr=0.72$.  Unlike~\cite{boin20063d}, the leading edge is not included in the
simulations. The inflow boundary is located at $x=0.3 x_{\text{sh}}$ where the
analytical compressible boundary layer solution of~\cite{white2006viscous} is
imposed. The session files used in this example can be found in
Example~\ref{f:suppl:swbli}.  At the inlet, the Rankine-Hugoniot relations that
describe the incident shock are superimposed over the compressible boundary
layer solution.  At the top boundary we impose the constant states corresponding
to inviscid post incident shock wave state.  At the outlet in the subsonic part
of the boundary layer a pressure outlet is imposed based on the inviscid post
reflected state conditions. All boundary conditions are imposed in a weak sense
through a Riemann solver, as described in~\cite{mengaldo2014guide}, and use a
coarse grid of $60 \times 40$ quadrilateral elements at order $p=3$. For
illustrative purposes, Figure~\ref{fig:app-aero-swbli-mach} shows a snapshot of
the Mach number field. For a more quantitative comparison,
Figure~\ref{fig:app-aero-swbli-cf} compares the skin friction coefficient with
those from~\cite{eckert1955engineering} and~\cite{boin20063d}, which is in fair
agreement with the results of \cite{boin20063d}.

\section{Availability}

\nek is open-source software, released under the MIT license, and is freely
available from the project website (\url{https://www.nektar.info/}). The {\tt
  git} repository is freely accessible and can be found at
\url{https://gitlab.nektar.info/}. Discrete releases are made at milestones in
the project and are available to download as compressed {\tt tar} archives, or
as binary packages for a range of operating systems. These releases are
considered to contain relatively complete functionality compared to the
repository {\tt master} branch. Docker container images are also available for
these releases and the latest build of {\tt master}, as well as a Jupyter
notebook that contains the Python interface of Section~\ref{sec:python}. These
can be found on Dockerhub under the repositories {\tt nektarpp/nektar} and {\tt
  nektarpp/nektar-workbook} respectively.

\section{Conclusions}
\label{sec:conclusions}

In this paper, we have reviewed the latest features and enhancements of the \nek
version 5.0 release. A key theme of our work in this release has been to evolve
the fundamental design of the software detailed in our previous
publication~\cite{nektarpp2015}, towards providing an enabling tool for
efficient high-fidelity simulations in various scientific areas. To this end,
this latest version of \nek provides a complete pipeline of tools: from
pre-processing with \emph{NekMesh} and a new parallel I/O interface for mesh and
field representations; new solvers and improvements to existing ones through
numerical developments such as spatially variable polynomial order and the
global mapping technique; to parallel post-processing and in-situ processing
with the \emph{FieldConvert} utility developments. This gives scientific
end-users a tool to enable efficient high-fidelity simulations in a number of
fields, such as the applications we discuss in Section~\ref{sec:applications}.

Although this version represents a major milestone in the development of \nek,
there is still clear scope for future work. A particular area of focus remains
the efficient use of many-core CPU and GPU systems, recognising that
optimisation and performance on an increasingly diverse range of hardware
presents a major challenge. Initial research in this area has investigated the
use of matrix-free methods as a potential route towards fully utilising
computational hardware even on unstructured grids, by combining efficient sum
factorisation techniques and the tensor-product basis for unstructured elements
presented in~\cite{ShKa95}. From the perspective of code maintainability, we
have also investigated various performance-portable programming models in the
context of mesh generation~\cite{eichstadt-2018} and implicit
solvers~\cite{eichstadt-2019}. Looking towards the next major release of \nek,
we envision the use of these studies as a guideline to implementing efficient
operators for the \shp element method, whilst retaining ease of use for the
development of increasingly efficient solvers.

\subsection*{Acknowledgements}



\noindent The development of \nek has been supported by a number of funding
agencies including the Engineering and Physical Sciences Research Council
(grants EP/R029423/1, EP/R029326/1 EP/L000407/1, EP/K037536/1, EP/K038788/1,
EP/L000261/1, EP/I037946/1, EP/H000208/1, EP/I030239/1, EP/H050507/1,
EP/D044073/1, EP/C539834/1), the British Heart Foundation (grants FS/11/22/28745
and RG/10/11/28457), the Royal Society of Engineering, European Union FP7 and
Horizon 2020 programmes (grant nos.~265780, 671571 and~675008), McLaren Racing,
the National Science Foundation (IIS-0914564, IIS-1212806 and DMS-1521748), the
Army Research Office (W911NF-15-1-0222), the Air Force Office of Scientific
Research and the Department of Energy.  HX acknowledges support from the NSF
Grants 91852106 and 91841303. SC acknowledges the support of the National
Research Foundation of Korea (No. 2016R1D1A1A02937255). KL acknowledges the
Seventh Framework Programme FP7 Grant No. 312444 and German Research Foundation
(DFG) Grant No. JA 544/37-2.



\appendix
\section{Supplementary material}

\begin{figure}[!ht]
  File: {\tt user-guide.pdf}
  \caption{User guide for \nek detailing compilation, installation, input
    format and usage, including examples.}
  \label{f:suppl:user-guide}
\end{figure}

\begin{figure}[!ht]
  File: {\tt insitu.zip}
  \caption{\nek input files for simulating flow past a cylinder at
    $Re = \numprint{80}$ using {\tt IncNavierStokesSolver}. This example uses
    the HDF5 input format described in Section~\ref{sec:parallel_io} and the
    in-situ processing facilities of Section~\ref{sec:insitu} to generate an
    animation of the vorticity field and show the von K\'arm\'an vortex shedding
    in this regime.}
  \label{f:suppl:cylinder}
\end{figure}

\begin{figure}[!ht]
  File: {\tt adaptiveOrder.zip}
  \caption{\nek input files for simulating flow over a NACA0012 wing at
    $Re = \numprint{50000}$ using {\tt IncNavierStokesSolver}. This example uses
    an adaptive-in-time polynomial order described in Section~\ref{sec:varp} to
    increase efficiency of the simulation when compared to a spatially-constant
    polynomial order.}
  \label{f:suppl:adaptive}
\end{figure}

\begin{figure}[!ht]
  File: {\tt wavyWing.zip}
  \caption{\nek input files for simulating flow over a wavy NACA0012 wing at
    $Re = \numprint{1000}$ using {\tt IncNavierStokesSolver}. This example uses
    the mapping technique described in Section~\ref{sec:mapping} to perform
    simulations in a quasi-3D setting.}
  \label{f:suppl:wavy}
\end{figure}

\begin{figure}[!ht]
  File: {\tt acoustic.zip}
  \caption{\nek input files for simulating the spinning vortex pair using the
    {\tt AcousticSolver} of Section~\ref{sec:acoustic} at a polynomial order of
    $P=5$, accelerated using the {\tt Collections} library described in
    Section~\ref{sec:collections}.}
  \label{f:suppl:acoustic}
\end{figure}

\begin{figure}[!ht]
  File: {\tt meshGen.zip}
  \caption{\emph{NekMesh} input files for a NACA0012 aerofoil section and T106C
    turbine blade geometry outlined in Section~\ref{sec:nekmesh}.}
  \label{f:suppl:meshgen}
\end{figure}

\begin{figure}[!ht]
  File: {\tt vivCylFlow.zip}
  \caption{\nek input files for simulating flow over a flexible cylinder
    $Re = \numprint{3900}$ using the {\tt IncNavierStokesSolver}. This example
    uses the the thick strip model outlined in Section~\ref{sec:viv} to reduce
    computational cost against a full 3D simulation.}
  \label{f:suppl:viv}
\end{figure}

\begin{figure}[!ht]
  File: {\tt shockBL.zip}
  \caption{\nek input files for simulating a shock boundary-layer interaction
    test case, at a Reynolds number $Re = 10^5$, Mach number
    $Ma = \numprint{2.15}$ and shock angle $\beta = \numprint{30.8}^\circ$ as
    outlined in Section~\ref{sec:aero}.}
  \label{f:suppl:swbli}
\end{figure}


\bibliographystyle{cpc-mod}
\bibliography{nektarpp_old,nektarpp_new}

\end{document}